\documentclass[12pt,reqno]{amsart}
\allowdisplaybreaks[4]
\usepackage{graphicx}
\usepackage{lipsum}
\usepackage{amssymb}
\usepackage{color}
\usepackage{amsmath}
\usepackage{cite}
\usepackage{subfigure}
\usepackage{graphicx}
\usepackage{epstopdf}
\usepackage{bibentry}
\usepackage{cases}
\newtheorem{theorem}{Theorem}

\newtheorem{remark}{Remark}

\begin{document}
\title{General higher-order breathers and rogue waves in the two-component
long-wave--short-wave resonance-interaction mode}
\author{Jiguang Rao$^{1}$,  Boris A. Malomed $^{2,3}$, Dumitru Mihalache $^{4}$,  Jingsong He$^{5,\ast}$}
\thanks{$^*$ Corresponding author: hejingsong@szu.edu.cn, jshe@ustc.edu.cn}
\dedicatory {$^{1}$School of Mathematics and Statistics, Hubei University of Science and Technology, Xianning, Hubei, 437100, P.\ R.\ China;\\
$^{2}$ Department of Physical Electronics, School of Electrical Engineering, Faculty of Engineering, and Center for Light--Matter Interaction, Tel Aviv University, P.O.B. 39040, Ramat Aviv, Tel Aviv, Israel; \\
$^{3}$ Instituto de Alta Investigaci{\'o}n, Universidad de Tarapac{\'a}, Casilla 7D, Arica, Chile;\\
$^{4}$ Horia Hulubei National Institute of Physics and Nuclear Engineering, P.O. Box MG--6, Magurele, RO-077125, Romania;\\
$^{5}$ Institute for Advanced Study, Shenzhen University, Shenzhen, Guangdong, 518060, P.\ R.\ China\\
}

\begin{abstract}
General higher-order breather and rogue wave (RW) solutions to the two-component long wave--short wave resonance
interaction (2-LSRI) model are derived via the bilinear Kadomtsev-Petviashvili hierarchy reduction method and
are given in terms of determinants. Under particular parametric conditions, the breather solutions can reduce
to homoclinic orbits, or a mixture of breathers and homoclinic orbits.  There are three families of RW solutions,
which correspond to a simple root, two simple roots,  and a double root of an algebraic equation related to the
dimension reduction procedure. The first family of RW solutions consists of $\frac{N(N+1)}{2}$ bounded fundamental RWs,
the second family is composed of $\frac{N_1(N_1+1)}{2}$ bounded fundamental RWs coexisting with another $\frac{N_2(N_2+1)}{2}$
fundamental RWs of different bounded state ($N,N_1,N_2$ being positive integers), while the third one have
${[\widehat{N}_1^2+\widehat{N}_2^2-\widehat{N}_1(\widehat{N}_2-1)]}$ fundamental bounded RWs ($\widehat{N}_1,\widehat{N}_2$ being non-negative integers).
The second family can be regarded as the superpositions of the first family, while the third family can be the degenerate case of the first family
under particular parameter choices. These diverse RW patterns are illustrated graphically. 
\end{abstract}
\maketitle
\section{Introduction}
Resonant three-wave interactions plays a crucially important role in various
setups which occur in Bose-Einstein condensates, fluid mechanics, optics,
and other areas of physics \cite{c-1,c-2,c-3,c-4,c-5}. Theoretical studies
of these phenomena provide an essential contribution to the nonlinear-wave
dynamics \cite{duction-2,inter-1,inter-2,inter-4,duction-1,inter-5}.
Generally, in nonlinear systems with linear dispersion relation $\omega
=\omega (k)$ ($\omega $ and $k$ are, as usual, the frequency and wavenumber,
respectively), the resonant interaction takes place if the corresponding
frequencies and wavenumbers, which form a \textit{resonance triad} \cite%
{LS-RI1}, are mutually locked by conditions%
\begin{equation}
k_{3}=k_{1}+k_{2},\omega _{3}=\omega _{1}+\omega _{2}.  \label{three waves}
\end{equation}%
The limit case of the triad amounts to the \textit{long-wave--short-wave}
(LW-SW) resonance, when one wave is much longer than the other two \cite%
{LS-RI1,LS-RI2,LS-RI3}, i.e.,
\begin{equation}
k_{3}=k+(\Delta {k})/2,k_{2}=k-(\Delta {k})/2,k_{1}=\Delta {k},|\Delta {k}%
|\ll |k|.
\end{equation}%
In this special case, condition (\ref{three waves}) for the frequency
amounts, in the first approximation, to%
\begin{equation}
{\mathrm{d}} {\omega }/{\mathrm{d}} {k}\approx \omega (\Delta {k})/\Delta {k},
\end{equation}%
i.e., the LW\ phase velocity must match the SW group velocity, at the
special value of wavenumber $k$. In this case, the nonlinear equation,
governing a slowly varying, complex-valued SW packet envelope ($S$) and a
real-valued LW field ($L$), has been derived by means of the multiple-scale
asymptotic expansion \cite{YO1,Ma-Yu1,Ma-Yu2}:
\begin{equation}
\begin{array}{c}
iS_{t}-S_{xx}+LS=0, \\
L_{t}=2(\delta |S|^{2})_{x}.%
\end{array}
\label{LSRI}
\end{equation}%
The sign of the real nonlinearity coefficient $\delta $ depends on the
physical realization of the system. It is often called the LW--SW resonant
interaction (LSRI) model, or the Yajima--Oikawa (YO) system. It applies to
fluid mechanics and a number of other physical settings.

The LSRI model \eqref{LSRI} is completely integrable as it admits a Lax
pair, and was solved by dint of the inverse scattering transform \cite{YO1}.
Additionally, Cheng had proposed the LSRI model \eqref{LSRI} from the
so-called $K$-constrained Kadomtsev-Petviashvili (KP) hierarchy with $K=2$,
while $K=1$ corresponds to the classical nonlinear Schr\"{o}dinger (NLS)
equation \cite{chen-JMP}. Bright- and dark-soliton solutions of Eq. %
\eqref{LSRI} were constructed in Ref. \cite{Ma-Yu1,Ma-Yu2}. The first-order
RW solutions to the LSRI model \eqref{LSRI} were studied by means of the
Hirota's bilinear method \cite{chow-jpsj} and Darboux transformation \cite%
{chen-PRE1,chen-pla}. The higher-order breather and RW solutions were
derived with the help pf the bilinear KP hierarchy reduction method \cite%
{chen-br,chen-jpsj}. Homoclinic connections of unstable plane waves were
investigated by means of the B\"{a}cklund transform \cite{Wright-1}.

RWs are large displacements from an otherwise tranquil (but usually
unstable) background \cite{rw-1,rw-2,rw-3,rw-4,rw-5,rw-6,rw-7,rw-8,rw-9},
whose most striking features are unpredictability and localization in time
and space \cite{ak-rw-1,ak-rw-2}. In the past decade, RWs have attracted a
great deal of interest \cite{xrw-1,xrw-2,xrw-3,xrw-4,xrw-5,xrw-6,xrw-8,xrw-9}
in the experimental and theoretical communities alike. Explicit solutions of
integrable equations for RWs help to understand these phenomena in the
physical systems. Such solutions have been found for many integrable models,
such as the NLS equation and its multicomponent version \cite%
{NLS-1,NLS-2,NLS-3,2NLS-1,2NLS-2,2NLS-3,2NLS-4,2NLS-5,2NLS-6,miller-1,miller-2}%
, the Davey--Stewartson (DS) \cite{Ohta-DS1,Ohta-DS2,jiguang-DSI} and
Ablowitz-Ladik \cite{abn-1,abn-2} equations, etc. \cite{orw-1,orw-2,orw-3,orw-4,orw-5,orw-6,orw-7,orw-8,orw-10,orw-11,orw-12}.
 The RW solutions can be regarded as the limit case of breathers, which are
periodic in time or spatial coordinate. The breathers can also provide a model for the observation of RWs in
experiments \cite{xrw-6}.  A rogue wave may be thought of as the consequence of modulation instability (MI), but
conversely not all of MIs necessarily result in rogue wave generation \cite{2NLS-4,MI-1}.  Besides, Baronio {\it et al.}
found that rogue wave solutions in several physical systems exist in the subset of parameters where the MI is present if and
only if the unstable sideband spectrum also contains continuous wave or zero-frequency perturbations as a limit case \cite{MI-2}.

In this paper, we consider the following two-component LSRI (2-LSRI) model
governing the resonance of two SW components with a common LW one:
\begin{gather}
iA_{t}-A_{xx}+LA=0,  \notag \\
iB_{t}-B_{xx}+LB=0,  \label{2-LSRI} \\
L_{t}=2(\delta _{1}|A|^{2}+\delta _{2}|B|^{2})_{x},  \notag
\end{gather}%
where $A$, $B$ are the SWs and $L$ is the LW, real nonlinear coefficients $%
\delta _{1},\delta _{2}$ depend on the precise physical properties of the
system, e.g. the density stratification profile in fluid \cite{c-1,Ma-Yu3}.
In optical contexts, the two short-wave components $A$ and $B$ are called optical
waves, while the long-wave component $L$ are regarded as
the induced optical rectification \cite{Duke} or the low--frequency
terahertz  wave \cite{A-PRL}.

The 2-LSRI model is an integrable extension of the LSRI model %
\eqref{LSRI}, and it reduces to the LSRI model \eqref{LSRI} for $A=B=S$.
This system also admits soliton solutions of bright, dark, and mixed
bright--dark types \cite{kanna-pre,Chen-so-1,Chen-so-2}. The
first-order RW solution was first constructed by the Darboux transformation
\cite{chen-PRE} and later by means of the Hirota's bilinear method \cite%
{Chow-Non}.
The connection between the existence criterion of
rogue waves and the onset of baseband MI in the LSRI model \eqref{LSRI} and the 2-LSRI model \eqref{2-LSRI} was confirmed in Ref. \cite{MI-2} and Ref.
\cite{Chow-Non}, respectively.
 Additionally, Chen {\it et.al.} have obtained several RW
solutions from the general rational solutions of the (2+1)-dimensional
multi-component LSRI system \cite{YO-PLA}. Very recently, Li and Geng have constructed a
family of higher-order bounded RW solutions consisting of $N(N+1)/2$
fundamental RWs with the help of the Darboux transformation \cite%
{geng-CNSNS,YO-Geng1,YO-Geng2}. That family of solutions of the 2-LSRI model %
\eqref{2-LSRI} shares the same RW patterns with the higher-order RW of LSRI model \cite{chen-jpsj},
and there not exist new RW patterns for 2-LSRI model in contrast with LSRI model.
It is known that the multi-component NLS equation admits more RW patterns in
contrast to the scalar NLS equation \cite{2NLS-1,2NLS-2,2NLS-3,2NLS-4,2NLS-5,2NLS-6}, such as the mixed
bounded RWs consisting of RWs of different types, and the degenerate bounded RWs.  Very recently,
Yang and Yang derived a family of new bounded
RW solutions  to the three-wave resonant interaction system \cite{orw-2}, which {consists
 of ${[\widehat{N}_1^2+\widehat{N}_2^2-\widehat{N}_1(\widehat{N}_2-1)]}$ fundamental
bounded RWs ($\widehat{N}_1,\widehat{N}_2$ being non-negative integers). By taking $\widehat{N}_1=0$ or
$\widehat{N}_2=0$,  this solution family amounts to one consisting of the degenerate bounded RW solutions. For
$\widehat{N}_1\widehat{N}_2\neq0$,  the family contains non-degenerate bounded RW solutions.}
Thus, we call this family of RW solutions as {\it degradable} bounded RW solutions in this paper.
Up to now, the higher-order mixed bounded and degradable bounded RWs were not
reported for the 2-LSRI model \eqref{2-LSRI}, to the best of our knowledge.
A natural motivation is to construct the higher-order mixed bounded and degradable bounded
RWs in the 2-LSRI model \eqref{2-LSRI}. Additionally, the first-order breather
solutions were constructed by means of the Hirota's bilinear method \cite%
{Chow-Non}, and the first-order homoclinic orbits were derived by the B\"{a}%
cklund transformation \cite{Wright-2}. However, higher-order breather and
homoclinic-orbit solutions have not been reported, as yet, for the 2-LSRI
model \eqref{2-LSRI}, to the best of our knowledge. Very recently, we
constructed general higher-order breather solutions for the multi-component
two-dimensional LSRI model in terms of determinants via the bilinear
KP-hierarchy reduction method \cite{RAO-2DLSRI}, which can also be applied
to derive the higher-order breather and homoclinic-orbit solutions of the
2-LSRI model \eqref{2-LSRI} in the form of determinants.

The main goal of the present paper is to construct the general higher-order
breather and RW solutions by employing the bilinear KP hierarchy reduction
method. The objectives of our work are as follows:

\begin{itemize}
\item The general higher-order breather solutions in terms of determinants
will be constructed. Under particular parametric restrictions, the breather
solutions can reduce to higher-order homoclinic orbits or a mixture of
homoclinic orbits and breathers.

\item Three families of RW solutions, corresponding to a simple root, two
simple roots, and a double root of an algebraic equation related to the
dimension reduction procedure, will be derived. The first family is the
bounded $N$-th-order RWs consisting of $N(N+1)/2$ fundamental RWs. The
second family is the mixed bounded $(N_{1},N_{2})$-th order RWs comprising $%
N_{1}(N_{1}+1)+N_{2}(N_{2}+1)/2$ fundamental RWs, in which the $N_{1}$%
-th-order bounded RWs and $N_{2}$-th bounded RW can be included in different
states. The third one is the degradable bounded $(\widehat{N}_{1},\widehat{N}%
_{2})$-th order RWs. Here $N,N_{1},N_{2}$ are positive integers, and $%
\widehat{N}_{1},\widehat{N}_{2}$ are non-negative integers. When $\widehat{N}%
_{1}=0$ or $\widehat{N}_{2}=0$, the third family solutions are degenerate
RWs.
\end{itemize}

The paper is organized as follows. In Section \ref{breathers}, we present
the general higher-order breather solutions for the 2-LSRI model in the form
of Theorem \ref{theorem1}, and then investigate dynamics of the breathers.
In Section \ref{3-RWS}, we present three different families of RW solutions
by three Theorems (i.e., Theorems \ref{theorem-RW-1}, \ref{theorem-RW-2}, %
\ref{theorem-RW-3}), then we exhibit dynamics of these three families of
RWs, respectively. In Section \ref{derivation}, we give the derivation of
the breather solutions in Theorem \ref{theorem1} and RW solutions in
Theorems \ref{theorem-RW-1}, \ref{theorem-RW-2}, \ref{theorem-RW-3}, i.e.,
the proofs of these four Theorems. The discussion of the obtained results
and the conclusions are presented in Section \ref{conclusion}.

\section{Dynamics of breathers in the 2-LSRI model }
\label{breathers}

This Section focuses on the dynamics of breathers in the 2-LSRI model %
\eqref{2-LSRI}. For this purpose, we first present the general breather
solutions in forms of determinants for the 2-LSRI model by the following
Theorem. The proof for this Theorem is postponed in Section \ref{derivation}.

\begin{theorem}\label{theorem1}
The 2-LSRI model \eqref{2-LSRI} admits the following breather solutions:
\begin{equation} \label{Bre-so}
\begin{aligned}
A=\rho_1e^{i\left(k_1x+(\gamma+k_1^2)t\right)}\frac{g}{f}, B=\rho_2e^{i\left(k_2x+(\gamma+k_2^2)t\right)}\frac{h}{f},L=\gamma-2( {\rm log} f)_{xx},
\end{aligned}
\end{equation}
 where the real function $f$ and the complex functions $g,h$  are given by
\begin{equation} \label{br-fg}
\begin{aligned}
f=\tau_{0,0},g=\tau_{1,0},h=\tau_{0,1}
\end{aligned}
\end{equation}
and  $\tau_{n,k}$ is defined as the following $N\times N$ determinant:
\begin{align}\label{br-tau}
&\tau_{n,k}=\det_{1 < s,j \leq N}\left(\overline{m}_{s,j}^{(n,k)}\right),
\end{align}
with
\begin{equation}\label{br-m}
\begin{aligned}
\overline{m}_{s,j}^{(n,k)}=&\frac{1}{p_s^{[1]}+p_j^{[1]*}}(-\frac{p_s^{[1]}-ik_1}{p_j^{[1]*}+ik_1})^n
(-\frac{p_s^{[1]}-ik_2}{p_j^{[1]*}+ik_2})^k
e^{\zeta_s+\zeta_j^{*}}+\frac{1}{p_s^{[1]}+p_j^{[2]*}}(-\frac{p_s^{[1]}-ik_1}{p_j^{[2]*}+ik_1})^n\\
&\times(-\frac{p_s^{[1]}-ik_2}{p_j^{[2]*}+ik_2})^k
e^{\zeta_s}+\frac{1}{p_s^{[2]}+p_j^{[1]*}}(-\frac{p_s^{[2]}-ik_1}{p_j^{[1]*}+ik_1})^n(-\frac{p_s^{[2]}-ik_2}{p_j^{[1]*}+ik_2})^k
e^{\zeta_j^*}+\\
&\frac{1}{p_s^{[2]}+p_j^{[2]*}}(-\frac{p_s^{[2]}-ik_1}{p_j^{[2]*}+ik_1})^n(-\frac{p_s^{[2]}-ik_2}{p_j^{[2]*}+ik_2})^k,
\end{aligned}
\end{equation} \\
and
\begin{equation}\label{br-xi}
\begin{aligned}
\zeta_s=&\xi_s^{[1]}-\xi_s^{[2]},\\
\xi_s^{[\alpha]}=&p_s^{[\alpha]}x-ip_s^{[\alpha]2}t+\overline{\xi}_s^{[\alpha]},\alpha=1,2.
 \end{aligned}
\end{equation}
In the above expressions, $p_s^{[\lambda]}, \overline{\xi}_s^{[\alpha]}$ are arbitrary complex constants,{$\rho_{\ell},k_{\ell},\gamma$ are freely real parameters,} and the parameters $p_s^{[1]},p_s^{[2]}$ have to satisfy the following constraints:
\begin{equation}\label{br-con}
\begin{aligned}
\frac{\delta_1\rho_1^2}{(p_s^{[1]}-ik_1)(p_s^{[2]}-ik_1)}+\frac{\delta_2\rho_2^2}{(p_s^{[1]}-ik_2)(p_s^{[2]}-ik_2)}-i(p_s^{[1]}+p_s^{[2]})=0.
 \end{aligned}
\end{equation}
\end{theorem}

\begin{remark}\label{remark1}
If we assume $p_{sR}^{[\alpha]}>0$, then $f$ in Eq. \eqref{br-fg} is positive, and the above breather solutions are nonsingular. Thus hereafter we take $p_{sR}^{[\alpha]}>0$ to avoid the singularities of the breathers.   The subscripts $R$ and $I$ represent the real and imaginary parts of a given parameter or a function, respectively.
\end{remark}

\begin{remark}\label{remark2}
When $\delta_1=\delta_2,k_1=k_2,\rho_1=k\rho_2$, the above breather solution reduces to the breather solutions of the scalar LSRI model \eqref{LSRI} reported in \cite{chen-br}. However, the breather solutions in \cite{chen-br} were given by $2N\times2N$ determinants, whereas they are given by $N\times N$ determinants in Theorem \ref{theorem1}.
\end{remark}

\begin{remark}\label{remark3}
If $p_{sR}^{[1]2}-p_{sI}^{[1]2}=p_{sR}^{[2]2}-p_{sI}^{[2]2}$, then the coefficients of $t$ in $\zeta_{sI}$ are zero, and the solutions \eqref{Bre-so} become the higher-order homclinic orbits. If $p_{sR}^{[1]2}-p_{sI}^{[1]2}=p_{sR}^{[2]2}-p_{sI}^{[2]2}, p_{jR}^{[1]2}-p_{jI}^{[1]2}\neq p_{jR}^{[2]2}-p_{jI}^{[2]2}$ ( $1\leq s\neq j \leq N$), the  solutions \eqref{Bre-so} are a mixture of breathers and homclinic orbits.
\end{remark}

By taking $N=1$ in Eq. \eqref{br-tau}, Theorem \ref{theorem1} yields the
first-order breather solutions of the 2-LSRI model, which can be expressed
in terms of hyperbolic and trigonometric functions as:
\begin{equation}  \label{br-1}
\begin{aligned}
A=&\overline{A}\frac{e^{\theta_0}\cosh(\zeta_{1R}+\theta_0+i\overline{%
\beta}_1)+e^{\widehat{\alpha}}\cos(\zeta_{1I}+\widehat{\beta}-i\overline{%
\alpha}_1)}{e^{\theta_0}\cosh(\zeta_{1R}+\theta_0)+e^{\widehat{\alpha}}\cos(%
\zeta_{1I}+\widehat{\beta})},\\
B=&\overline{B}\frac{e^{\theta_0}\cosh(\zeta_{1R}+\theta_0+i\overline{%
\beta}_2)+e^{\widehat{\alpha}}\cos(\zeta_{1I}+\widehat{\beta}-i\overline{%
\alpha}_2)}
{e^{\theta_0}\cosh(\zeta_{1R}+\theta_0)+e^{\widehat{\alpha}}\cos(\zeta_{1I}+%
\widehat{\beta})},\\
L=&\gamma-\frac{y_2\cosh(\zeta_{1R}+\theta_0)\cos(\zeta_{1I}+\widehat{%
\beta})+y_1\sinh(\zeta_{1R}+\theta_0)\sin(\zeta_{1I}+\widehat{\beta})+y_0}{%
\left(e^{\theta_0}
\cosh(\zeta_{1R}+\theta_0)+e^{\widehat{\alpha}}\cos(\zeta_{1I}+\widehat{%
\beta})\right)^2}. \end{aligned}
\end{equation}
The auxiliary functions in the above expressions are defined by
\begin{equation}
\begin{aligned}
&\zeta_1=\xi_1^{[1]}-%
\xi_1^{[2]}=(p_1^{[1]}-p_1^{[2]})x-i(p_1^{[1]2}-p_1^{[2]2})t+\overline{%
\xi}_1^{[1]}-\overline{\xi}_1^{[2]},\\
&\overline{A}=-\rho_1(p_1^{[2]}+p_1^{[2]*})%
\frac{p_1^{[2]}+ik_1}{p_1^{[2]}-ik_1}e^{i\left(k_1x+(\gamma+k_1^2)t+%
\overline{\beta}_1\right)},\\
&\overline{B}=-\rho_2(p_1^{[2]}+p_1^{[2]*})%
\frac{p_1^{[2]}+ik_2}{p_1^{[2]}-ik_2}e^{i\left(k_2x+(\gamma+k_2^2)t+%
\overline{\beta}_2\right)},\\
&e^{\overline{\alpha}_\ell+i\overline{\beta}_\ell}=\frac{p_1^{[1]}-ik_%
\ell}{p_1^{[2]}-ik_{\ell}},e^{\widehat{\alpha}+i\widehat{\beta}}=%
\frac{p_1^{[2]}+p_1^{[2]}}{p_1^{[2]}+p_1^{[1]}},\ell=1,2,\\
&y_1=2e^{\theta_0+\widehat{\alpha}}(\overline{c}^2-\overline{e}^2),y_2=4e^{%
\theta_0+\widehat{\alpha}}\overline{c}\overline{e},y_0=2(\overline{c}^2e^{2%
\theta_0}-\overline{e}^2e^{2\widehat{\alpha}}),\\
&\overline{c}=\frac{1}{2}\left(p_1-p_2+p_1^*-p_2^*\right),\overline{e}=%
\frac{1}{2i}\left(p_1-p_2-(p_1^*-p_2^*)\right),e^{\theta_0}=\sqrt{%
\frac{p_2+p_2^*}{p_1+p_1^*}}, \end{aligned}
\end{equation}
and $\xi_1^{[1]},\xi_1^{[2]}$ are defined in Eq. \eqref{br-xi}.

\begin{figure}[!htb]
\centering
\subfigure{\includegraphics[height=3.0cm,width=14cm]{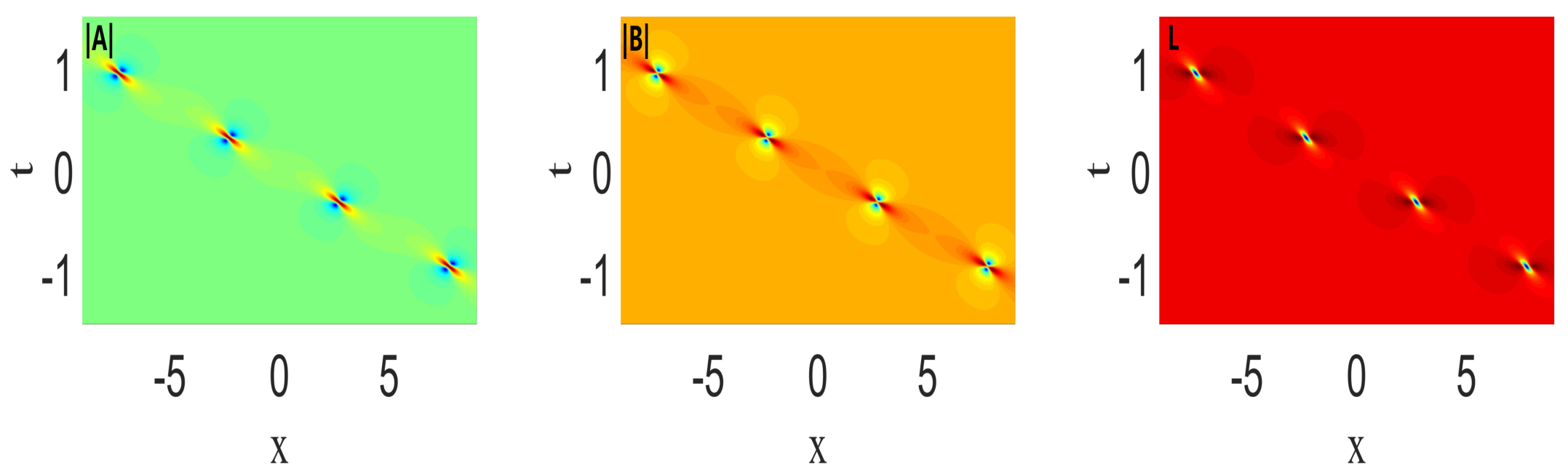}}
\caption{(Colour online) The first-order breather in the 2-LSRI model
\eqref{2-LSRI}, which is described by the solutions \eqref{br-1} with
parameters $\protect\delta_1=-1,\protect\delta_2=1,\protect\rho_1=\frac{%
\protect\sqrt{195}}{2},\protect\rho_2=\protect\sqrt{3},k_1=0,k_2=-1,\protect%
\gamma=0,p_1^{[1]}=2+i,p_1^{[2]}=3-2i,\overline{\protect\xi}_1^{[1]}=0,%
\overline{\protect\xi}_1^{[2]}=0$.~}
\label{fig1}
\end{figure}

From the above expressions of the first-order breather solution \eqref{br-1}%
, we can obtain that the first-order breather propagates along the line $%
\zeta _{1R}+\theta _{0}=0$, and is periodic along the line $\zeta _{1I}+%
\widehat{\beta }=0$. Figure \ref{fig1} shows the first-order breather in the
2-LSRI model \eqref{2-LSRI}.

\begin{figure}[!htb]
\centering
\subfigure{\includegraphics[height=3.5cm,width=14cm]{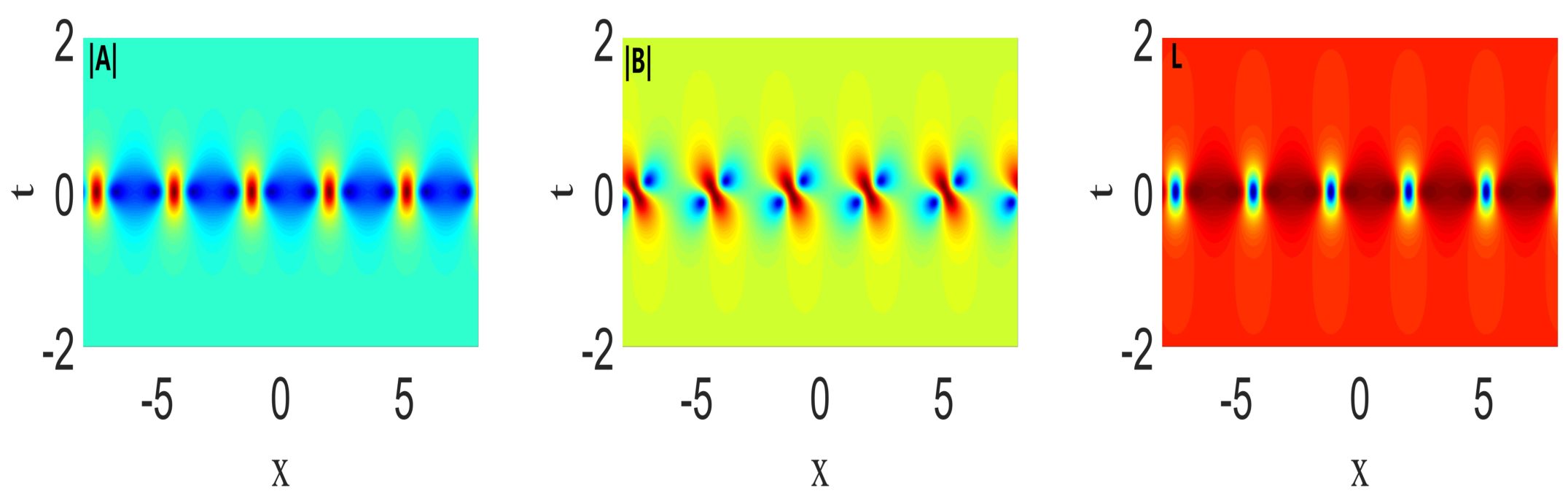}}
\caption{(Colour online) The first-order homoclinic orbit in the 2-LSRI
model \eqref{2-LSRI}, which is described by the solutions \eqref{br-1} with
parameters $\protect\delta_1=1,\protect\delta_2=-1,\protect\rho_1=\protect%
\sqrt{2},\protect\rho_2=\protect\sqrt{5},k_1=0,k_2=-1,\protect\gamma%
=0,p_1^{[1]}=1+i,p_1^{[2]}=1-i,\overline{\protect\xi}_1^{[1]}=0,\overline{%
\protect\xi}_1^{[2]}=0$.~}
\label{fig2}
\end{figure}

When $p_{1R}^{[1]2}-p_{1I}^{[1]2}=p_{1R}^{[2]2}-p_{1I}^{[2]2}$, the
coefficients of $t$ in $\zeta _{1I}$ are zero, hence the first-order
breather solutions \eqref{br-1} reduce to the first-order homoclinic orbit
solutions. Figure \ref{fig2} shows the first-order homoclinic orbit in the
2-LSRI model \eqref{2-LSRI} with parameters $p_{1}^{[1]}=1+i,p_{1}^{[2]}=1-i$%
.

Here we have to note that the first-order breather solutions \eqref{br-1}
for the 2-LSRI model \eqref{2-LSRI} have been reported in \cite{Chow-Non},
and were derived by the Hirota's direction method. By taking the long wave
limit of the obtained first-order breather solution, the first-order RW solutions
 were also derived for the 2-LSRI model \eqref{2-LSRI}.
 Additionally, as shown in Figs. \ref{fig1},\ref{fig2}, the
LW component $L$ has wave structures similar to those of the SW components $%
A $ and $B$, so we only focus on the dynamics of breathers in the SW
components $A$ and $B$, and in what follows we will not show the LW
component.


\begin{figure}[!htb]
\centering
\subfigure{\includegraphics[height=7cm,width=14cm]{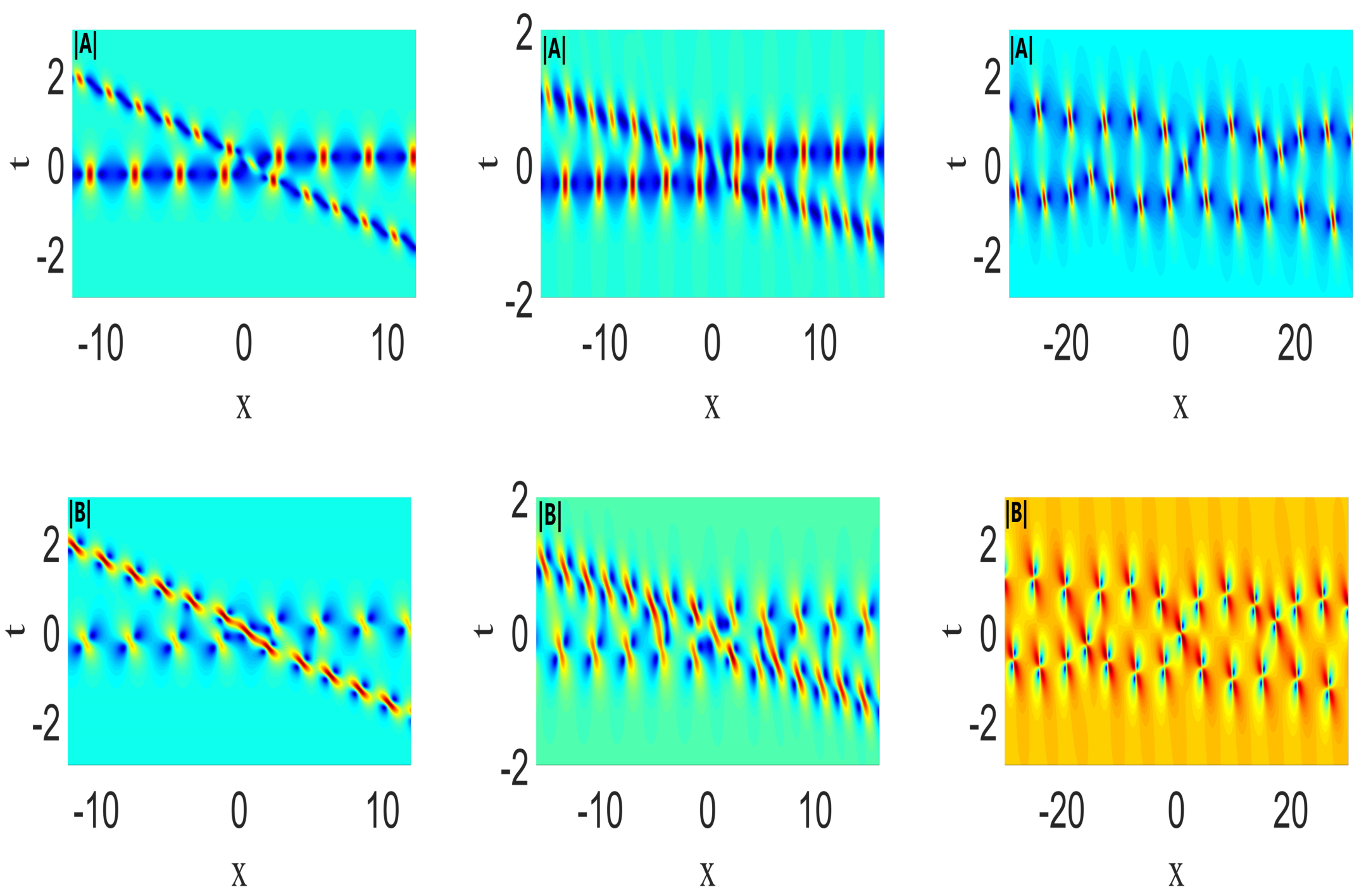}}
\caption{(Colour online) The leftmost panels: the second-order homoclinic
orbits with parameters $\protect\delta_1=1,\protect\delta_2=-1,\protect\rho%
_1=\protect\sqrt{2},\protect\rho_2=\protect\sqrt{5},k_1=0,k_2=-1,\protect%
\gamma%
=0,p_1^{[1]}=1+i,p_1^{[2]}=1-i,p_2^{[1]}=1.78521+2i,p_2^{[2]}=0.53496-1.048432i
$; The middle panels: a mixture of a first-order breather and a first-order
homoclinic orbit with parameters $\protect\delta_1=1,\protect\delta_2=-1,%
\protect\rho_1=\protect\sqrt{2},\protect\rho_2=\protect\sqrt{5},k_1=0,k_2=-1,%
\protect\gamma%
=0,p_1^{[1]}=1+i,p_1^{[2]}=1-i,p_2^{[1]}=1+2i,p_2^{[2]}=0.6780-1.2539i$; The
rightmost panels: the second-order breather with parameters $\protect\delta%
_1=-1,\protect\delta_2=-1,\protect\rho_1=\protect\sqrt{\frac{5}{6}},\protect%
\rho_2=\protect\sqrt{\frac{55}{12}},k_1=0,k_2=-1,\protect\gamma%
=0,p_1^{[1]}=1+i,p_1^{[2]}=1-\frac{1}{2}i,p_2^{[1]}=1-\frac{1}{3}%
i,p_2^{[2]}=1.0362+0.7905i$. ~}
\label{fig3}
\end{figure}

By taking $N=2$ in Eq. \eqref{br-tau}, Theorem \ref{theorem1} generates the
second-order breather solutions of the 2-LSRI model. The determinant forms
of the functions $f,g$, and $h$ of solutions \eqref{Bre-so} can be
explicitly written as
\begin{equation}
\begin{aligned} f=&\begin{vmatrix}
\overline{m}_{1,1}^{(0,0)}&\overline{m}_{1,2}^{(0,0)}\\
\overline{m}_{2,1}^{(0,0)}&\overline{m}_{2,2}^{(0,0)} \end{vmatrix}, \quad
g=&\begin{vmatrix} \overline{m}_{1,1}^{(1,0)}&\overline{m}_{1,2}^{(1,0)}\\
\overline{m}_{2,1}^{(1,0)}&\overline{m}_{2,2}^{(1,0)} \end{vmatrix},\quad
h=&\begin{vmatrix} \overline{m}_{1,1}^{(0,1)}&\overline{m}_{1,2}^{(0,1)}\\
\overline{m}_{2,1}^{(0,1)}&\overline{m}_{2,2}^{(0,1)} \end{vmatrix},
\end{aligned}  \label{2-fg}
\end{equation}%
where $\overline{m}_{s,j}^{(n,k)}$ are given by Eq. \eqref{br-m}. Since this
second-order breather is a superposition of first-order breathers, thus
these solutions have three different dynamical behaviours: (i) the
second-order homoclinic orbits for $%
p_{sR}^{[1]2}-p_{sI}^{[1]2}=p_{sR}^{[2]2}-p_{sI}^{[2]2},s=1,2$; (ii) the
mixture of the first-order breather and first-order homoclinic orbit for $%
p_{sR}^{[1]2}-p_{sI}^{[1]2}=p_{sR}^{[2]2}-p_{sI}^{[2]2}$ and $%
p_{3-s\,R}^{[1]2}-p_{3-s\,I}^{[1]2}\neq \,p_{3-s\,R}^{[2]2}-p_{3-s\,I}^{[2]2}
$; and (iii) the second-order breather for $p_{sR}^{[1]2}-p_{sI}^{[1]2}\neq
p_{sR}^{[2]2}-p_{sI}^{[2]2}$. Figure \ref{fig3} shows these three different
types of periodic waves of the 2-LSRI model \eqref{2-LSRI}. In the leftmost
panels of Fig. \ref{fig3}, corresponding to the second-order homoclinic
orbits, the two periodic waves are only periodic in space (i.e., $x$); in
the middle panels, corresponding to a mixture of a first-order breather and
a first-order homclinic orbit, one periodic wave (namely, the one
propagating along the $x$-axis) is only periodic along space $x$ and the
other periodic wave is periodic along both space $x$ and time $t$; in the
rightmost panels, both of the two periodic waves are periodic along space $x$
and time $t$.

For larger $N$ in Theorem \ref{theorem1}, the higher-order breathers, or
higher-order homclinic orbits, or a mixture of them to the 2-LSRI model can
be obtained, namely a superposition of $N$ individual first-order solutions
given by Eq. \eqref{br-1}.

\section{Dynamics of rogue waves in the 2-LSRI model}
\label{3-RWS}
In this Section, we study the dynamics of RWs in the 2-LSRI
model \eqref{2-LSRI}. The RW solutions, which will be explicitly expressed
in this Section, depend on the root structure of the following algebraic
equation
\begin{equation}  \label{root}
\begin{aligned} \frac{\partial \mathcal{Q}}{\partial p}=0, \end{aligned}
\end{equation}
where
\begin{equation}  \label{root-Q}
\begin{aligned}
\mathcal{Q}(p)=\frac{\delta_1\rho_1^2}{p-ik_1}+\frac{\delta_2%
\rho_2^2}{p-ik_2}+ip^2. \end{aligned}
\end{equation}
For different types of roots of Eq. \eqref{root}, the algebraic expressions
of the corresponding RW solutions to the 2-LSRI model \eqref{2-LSRI} are
different. In what follows, we will mainly discuss the RW solutions
associated to a non-imaginary simple root, two non-imaginary simple roots,
and a non-imaginary double root of Eq. \eqref{root}, respectively. These
three families of solutions are bounded RWs, mixed bounded RWs, and the
degradable bounded RWs.

\begin{remark}\label{remark-rw1}
If $p_0$ is a root of Eq. \eqref{root}, then $-p_0^*$ is also a root for Eq. \eqref{root}, thus hereafter we assume $p_{0R}>0$ without loss of generality.
\end{remark}

\begin{remark}\label{remark-sec2-1}
Eq. \eqref{root} is a quintic equation having up to five roots. If $p_0$  is a triple root of Eq. \eqref{root}, then $-p_0^*$ is also a triple root of Eq. \eqref{root}, thus $p_0=-p_0^*$ must hold,  otherwise there are six roots for Eq. \eqref{root}, so $p_0$ is pure imaginary when it is a triple root of Eq. \eqref{root}.  Besides, there is a factor $\frac{1}{p_0+p_0^*}$ in RW solutions, hence $p_0$ cannot be pure imaginary. Thus, there do not exist RW solutions when $p_0$ is a triple root of Eq. \eqref{root}.  That is also true for  $p_0$ being a quadruple or a quintuple root of Eq. \eqref{root}, so we only consider the RW solutions associated with $p_0$ being a non-imaginary simple root, two non-imaginary simple roots, and a non-imaginary double root of Eq. \eqref{root}.
\end{remark}

Below, we give three families of higher-order RW solutions to the 2-LSRI
model \eqref{2-LSRI} and then consider their dynamics. These RW solutions
are expressed through both differential operator forms and Schur
polynomials. Before that, we have to review the definition of these Schur
polynomials $\mathbf{S}_j(\mathbf{x})$ with $\mathbf{x}=(x_1,x_2,\cdots)$:
\begin{equation}  \label{Sch-1}
\begin{aligned}
\sum\limits_{j=0}^{\infty}S_j(\mathbf{x})\epsilon^j=\exp(\sum\limits_{k=1}^{%
\infty}x_k\epsilon^k), \end{aligned}
\end{equation}
or more explicitly
\begin{equation}  \label{Sch-2}
\begin{aligned} &S_0(\mathbf{x})=1,\\ &S_1(\mathbf{x})=x_1,\\
&S_2(\mathbf{x})=\frac{1}{2}x_1^2+x_2,\\ &\vdots \\
&S_j(\mathbf{x})=\sum\limits_{l_1+2l_2+\cdots
ml_m=j}\left(\prod\limits_{k=1}^m\frac{x_k^{l_k}}{l_k!}\right). \end{aligned}
\end{equation}

\subsection{The bounded $N$-th-order rogue wave solutions and their dynamics}

\label{rw1-1}

\subsubsection{The bounded $N$-th-order rogue wave solutions}

If $p_{0}$ is a non-imaginary simple root of Eq. \eqref{root}, the bounded $N
$-th-order RW solutions of the 2-LSRI model are expressed as the following
Theorem.
\begin{theorem}\label{theorem-RW-1}
The 2-LSRI model \eqref{2-LSRI} admits the following bounded $N$th-order RW solutions
\begin{equation} \label{RW-1}
\begin{aligned}
A=\rho_1e^{i\left(k_1x+(\gamma+k_1^2)t\right)}\frac{g}{f}, B=\rho_2e^{i\left(k_2x+(\gamma+k_2^2)t\right)}\frac{h}{f},L=\gamma-2( {\rm log} f)_{xx},
\end{aligned}
\end{equation}
 where
\begin{equation} \label{rw1-fg}
\begin{aligned}
f=\tau_{0,0},g=\tau_{1,0},h=\tau_{0,1}
\end{aligned}
\end{equation}
and  $\tau_{n,k}$ is defined as the following $N\times N$ determinant:
\begin{align}\label{rw1-tau}
&\tau_{n,k}=\det_{1 < s,j \leq N}\left({m}_{2s-1,2j-1}^{(n,k)}\right),
\end{align}
and the matrix elements ${m}_{s,j}^{(n,k)}$ are given either
(a) in differential operator form:
\begin{equation}\label{rw1-m}
\begin{aligned}
 {m}_{s,j}^{(n,k)}=&\left.\frac{[\mathcal{T}(p)\partial_{p}]^s}{s!}\frac{[\mathcal{T}^*(p)\partial_{p^*}]^j}{j!}
m \right|_{p=p_0},\\
m=&\frac{1}{p+p^*}(-\frac{p-ik_1}{p^*+ik_1})^n(-\frac{p-ik_2}{p^*+ik_2})^ke^{\xi+\xi^*},
\end{aligned}
\end{equation}
or (b) through Schur polynomials:
\begin{equation}\label{rw1-cm}
\begin{aligned}
\widehat{m}_{s,j}^{(n,k)}=&\sum\limits_{\nu=0}^{\mathrm{min}(i,j)}[\frac{|\lambda_1|^2}{(p_0+p_0^*)^2}]^{\nu}\mathbf{S}_{i-\nu}({ \mathbf{x}}^{+}(n,k)+\nu{\bf s})
 \mathbf{S}_{j-\nu}({ \mathbf{x}}^{-}(n,k)+\nu{\bf s}^*).
\end{aligned}
\end{equation}
\noindent Here $N$ is an arbitrary positive integer.
The auxiliary functions $\mathcal{T}(p),\xi$ in Eq. \eqref{rw1-m} are defined by
\begin{equation}\label{rw1-TQ}
\begin{aligned}
\xi=&px-ip^2t+\sum\limits_{r=1}^{\infty}\widehat{a}_r\ln^r\mathcal{W}(p),\\
\mathcal{T}(p)=&\sqrt{\frac{\mathcal{Q}^2(p)-\mathcal{Q}^2(p_0)}{\mathcal{Q}^{'2}(p)}},\\
\mathcal{W}(p)=&\frac{\mathcal{Q}(p)\pm\sqrt{\mathcal{Q}^2(p)-\mathcal{Q}(p_0)}}{\mathcal{Q}(p_0)}.
\end{aligned}
\end{equation}
The vectors $\mathbf{x}^{\pm}(n,k)=(x_1^{\pm},x_2^{\pm},\cdots)$ in Eq. \eqref{rw1-cm} are defined as:
\begin{equation}\label{rw1-x}
\begin{aligned}
x_r^{+}(n,k)=&\lambda_rx-{i}\beta_rt+n\theta_r^{(1)}+k\theta_r^{(2)}+{a}_r,  \\
x_r^{-}(n,k)=&\lambda_r^*x+{i}\beta_r^*t-n\theta_r^{(1)}-k\theta_r^{(2)}+{a}_r^*,
\end{aligned}
\end{equation}
where $\alpha_r,\beta_r,\theta_r^{(1)}$, and $\theta_r^{(2)}$ are coefficients from the expansions
\begin{equation}\label{rw1-abl}
\begin{aligned}
p(\kappa)-p_0=&\sum\limits_{r=1}^{\infty}\lambda_r\kappa^{r},\\
p^2(\kappa)-p^2_0=&\sum\limits_{r=1}^{\infty}\beta_r\kappa^{r},\\
\ln\frac{p(\kappa)-ik_1}{p_0-ik_1}=&\sum\limits_{r=1}^{\infty}\theta_r^{(1)}\kappa^{r},\\
\ln\frac{p(\kappa)-ik_2}{p_0-ik_2}=&\sum\limits_{r=1}^{\infty}\theta_r^{(2)}\kappa^{r},
\end{aligned}
\end{equation}
and the vector $\mathbf{s}=(s_1,s_2,s_3,\cdots)$ is defined by the expansion:
\begin{equation}\label{rw1-s}
\begin{aligned}
\ln\left[\frac{1}{\kappa}(\frac{p_0+p_0^*}{p_1})(\frac{p(\kappa)-p_0}{p(\kappa)+p_0^*})\right]=\sum\limits_{r=1}^{\infty}s_r\kappa^{r}.
\end{aligned}
\end{equation}

\end{theorem}

\subsubsection{Dynamics of the bounded rogue waves}

We first consider the fundamental (i.e., first-order) RW solutions of the
2-LSRI model \eqref{2-LSRI}, which can be regarded as the basic elements of
high-order bounded RW solutions. For this purpose, we take $N=1$ in Theorem %
\ref{theorem-RW-1}, the first-order RW solutions of the 2-LSRI model %
\eqref{2-LSRI} are explicitly expressed as
\begin{equation}  \label{1RW-1}
\begin{aligned} A=\rho_1e^{i\left(k_1x+(\gamma+k_1^2)t\right)}\frac{g}{f},
B=\rho_2e^{i\left(k_2x+(\gamma+k_2^2)t\right)}\frac{h}{f},L=\gamma-2( {\rm
log} f)_{xx}, \end{aligned}
\end{equation}
where
\begin{equation}  \label{1RW-1fg}
\begin{aligned}
f=&\widehat{m}_{1,1}^{(0,0)},g=\widehat{m}_{1,1}^{(1,0)},h=%
\widehat{m}_{1,1}^{(0,1)},\\
\widehat{m}_{1,1}^{(n,k)}=&(\lambda_1x-{i}\beta_1t+n\theta_1^{(1)}+k%
\theta_1^{(1)})(\lambda_1^*x+i\beta_1^*t-n\theta_1^{(1)}-k\theta_1^{(2)})+%
\zeta_0, \end{aligned}
\end{equation}
and
\begin{equation}  \label{RW-xix}
\begin{aligned}
&\lambda_1=\frac{dp(\kappa)}{d\kappa}|_{\kappa=0},\beta_1=2\lambda_1p_0,\\
&\theta_1^{(1)}=\frac{\lambda_1}{p_0-{i}k_1},\theta_1^{(2)}=\frac{%
\lambda_1}{p_0-{i}k_2},\zeta_0=\frac{|\lambda_1|^2}{(p_0+p_0^*)^2},
\end{aligned}
\end{equation}
Here we have taken $a_1=0$ so that the center of the fundamental RW is
located at $x=0,t=0$. After simple algebraic calculations, the above RW
solutions can also be expressed as follows:
\begin{equation}  \label{1RW-1-Final}
\begin{aligned}
A=&\widehat{\rho}_1\left[1-\frac{2i(\widehat{a}_1\ell_2-\widehat{b}_1%
\ell_1)+(\widehat{a}_1^2+\widehat{b}_1^2)}{\ell_1^2+\ell_2^2+\zeta_0}%
\right], \\
B=&\widehat{\rho}_2\left[1-\frac{2i(\widehat{a}_2\ell_2-\widehat{b}_2%
\ell_1)+(\widehat{a}_2^2+\widehat{b}_2^2)}{\ell_1^2+\ell_2^2+\zeta_0}%
\right], \\
L=&\gamma+4\frac{\ell_1^2-\ell_2^2-\zeta_0}{(\ell_1^2+\ell_2^2+\zeta_0)^2},
\end{aligned}
\end{equation}
where $\widehat{\rho}_j=\rho_je^{i\left(k_jx+(\gamma+k_j^2)t\right)},%
\ell_1=x+2p_{0I}t,\ell_2=-2p_{0R}t,\zeta_0=\frac{1}{4p_{0R}^2},\widehat{a}_j=%
\frac{p_{0R}}{p_{0R}^2+(p_{0I}-k_j)^2}, \widehat{b}_j=\frac{k_j-p_{0I}}{%
p_{0R}^2+(p_{0I}-k_j)^2}$ for $j=1,2$.

This fundamental RW in the $A$ component is classified into three different
types:

\begin{itemize}
\item a bright RW when $(p_{0I}-k_1)^2\leq\frac{1}{3}p_{0R}^2$;

\item a four--petaled RW when $\frac{1}{3}p_{0R}^2<(p_{0I}-k_1)^2<3p_{0R}^2$;

\item a dark RW when $(p_{0I}-k_1)^2\geq3p_{0R}^2$.
\end{itemize}

\begin{figure}[!htb]
\centering
\subfigure{\includegraphics[height=5cm,width=14cm]{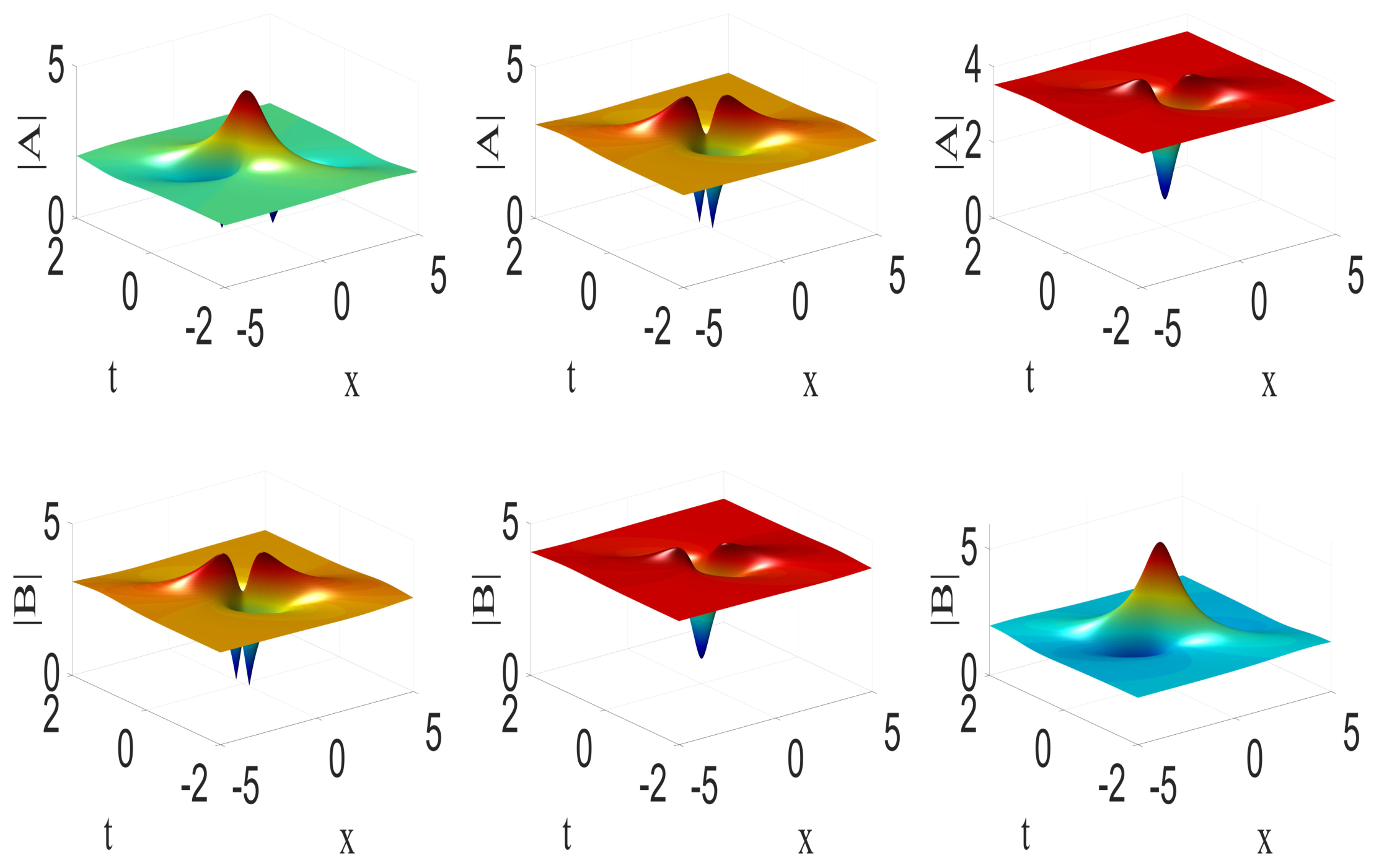}}
\caption{(Colour online) The three different types of first-order RW solutions \eqref{1RW-1-Final}
in the LSRI model. The leftmost panels: $\protect\delta_1=-1,\protect\delta_2=-1,\protect\rho_1=%
\frac{5\protect\sqrt{6}}{6},\protect\rho_2=\frac{2\protect\sqrt{21}}{3},k_1=%
\frac{3}{2},k_2=0,\protect\gamma=0,p_1^{[1]}=1+i$; The middle panels: $%
\protect\delta_1=-1,\protect\delta_2=1,\protect\rho_1=\frac{2\protect\sqrt{21%
}}{3},\protect\rho_2=\frac{5\protect\sqrt{6}}{3},k_1=0,k_2=-1 $; The
rightmost panels: $\protect\delta_1=-1,\protect\delta_2=-1,\protect\rho_1=%
\frac{5\protect\sqrt{2}}{2},\protect\rho_2=\frac{\protect\sqrt{14}}{2}%
,k_1=-1,k_2=1,\protect\gamma=0,p_1^{[1]}=1+i$. ~}
\label{fig4}
\end{figure}

By replacing $k_1$ with $k_2$ in the corresponding parameter condition, this
classification is also valid for the fundamental RW in the $B$ component.
The LW component $L$ is always a dark RW. Fig. \ref{fig4} displays these
three different types of fundamental RWs in both $A$ and $B$ components.

These fundamental RWs admit the center amplitudes given below:
\begin{equation}  \label{Center-AS1}
\begin{aligned}
|A^{c}|=&|\rho_1|\left|1-\frac{4p_{0R}^2}{p_{0R}^2+(p_{0I}-k_1)^2}\right|,
\\
|B^{c}|=&|\rho_2|\left|1-\frac{4p_{0R}^2}{p_{0R}^2+(p_{0I}-k_2)^2}\right|,
\\ L^{c}=&\gamma-16p_{0R}^2, \end{aligned}
\end{equation}
which are calculated at the origin \cite{orw-10,orw-11}. The
peak-to-background ratios of the fundamental RWs in the two SW components
are $|A^{(c)}|/|\rho_1|,|B^{(c)}|/|\rho_2|\leq3$, which indicates that the
fundamental RWs cannot reach a peak amplitude that exceeds three times the
background level, as in the case of fundamental RWs in the 2-NLS equations
\cite{orw-10,orw-11}.

\begin{figure}[!htb]
\centering
\subfigure{\includegraphics[height=5cm,width=14cm]{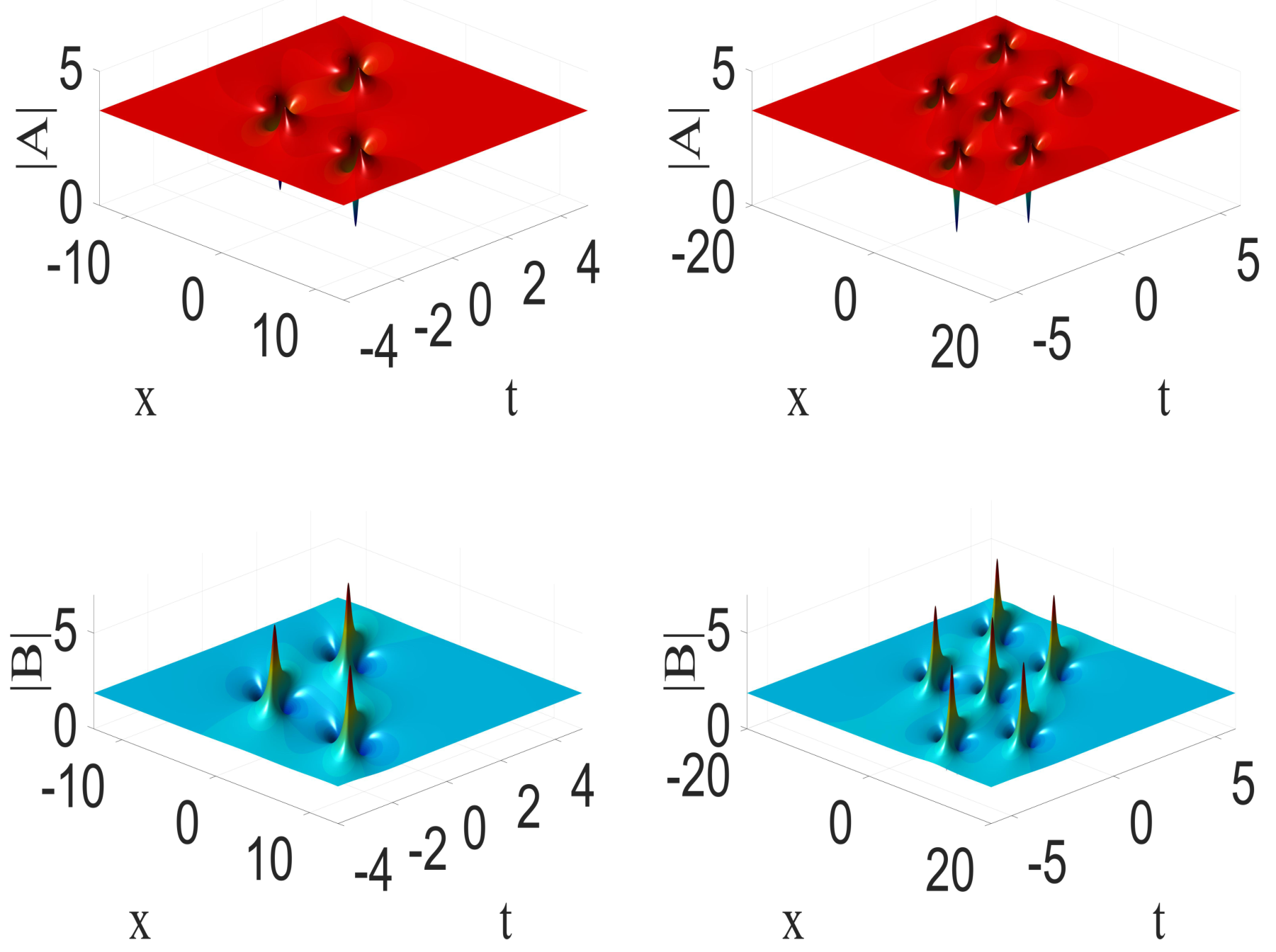}}
\caption{(Colour online) The left panels: the second-order bounded RW
solutions \eqref{RW-1} with parameters $N=2,\protect\delta_1=-1,\protect%
\delta_2=-1,\protect\rho_1=\frac{5\protect\sqrt{2}}{2},\protect\rho_2=\frac{%
\protect\sqrt{14}}{2},k_1=-1,k_2=1,\protect\gamma%
=0,p_1^{[1]}=1+i,a_1=0,a_2=0,a_3=100i$. The right panels: the third-order
bounded RW solutions \eqref{RW-1} with parameters $N=3,\protect\delta_1=-1,%
\protect\delta_2=-1,\protect\rho_1=\frac{5\protect\sqrt{2}}{2},\protect\rho%
_2=\frac{\protect\sqrt{14}}{2},k_1=-1,k_2=1,\protect\gamma%
=0,p_1^{[1]}=1+i,a_1=0,a_2=0,a_3=0,a_4=0,a_5=1000-5000i$. ~}
\label{fig5}
\end{figure}

The higher-order bounded RW solutions in Theorem \ref{theorem-RW-1} are
superpositions of $N(N+1)/2$ ($N\geq2$) fundamental RWs \eqref{1RW-1}. The
coexistence of these $N(N+1)/2$ fundamental RWs can generate diverse
waveforms. For instance, taking $N=2$ and $N=3$ in Theorem \ref{theorem-RW-1}%
, the second-order and third-order RW solutions can be derived,
respectively. These second-order and third-order RWs are displayed in Fig. %
\ref{fig5}. It is seen that the second-order RWs are composed of three
fundamental RWs (the left panels of Fig.\ref{fig5}), which form triangle patterns. The third-order RWs
comprising six fundamental RWs exhibit ring waveforms (the right panels of Fig.\ref{fig5}).


\subsection{The mixed bounded $(N_{1},N_{2})$-th order rogue wave solutions
and their dynamics}

\label{rw1-2}

\subsubsection{The mixed bounded $(N_{1},N_{2})$-th order rogue wave
solutions}

If $p_{0}^{(1)}$ and $p_0^{(2)}$ ($p_0^{(1)}\neq p_0^{(2)}$) are two
non-imaginary simple roots of Eq. \eqref{root}, the mixed RW solutions
comprising of bounded $N_1$th-order RWs and $N_2$th-order RWs are given by
the following Theorem.
\begin{theorem}\label{theorem-RW-2}
The 2-LSRI model \eqref{2-LSRI} admits the following mixed bounded $(N_1,N_2)$th-order RW solutions
\begin{equation} \label{RW-2}
\begin{aligned}
A=\rho_1e^{i\left(k_1x+(\gamma+k_1^2)t\right)}\frac{g}{f}, B=\rho_2e^{i\left(k_2x+(\gamma+k_2^2)t\right)}\frac{h}{f},L=\gamma-2( {\rm log} f)_{xx},
\end{aligned}
\end{equation}
 where
\begin{equation} \label{rw2-fg}
\begin{aligned}
f=\tau_{0,0},g=\tau_{1,0},h=\tau_{0,1}
\end{aligned}
\end{equation}
and  $\tau_{n,k}$ is defined as the following $2\times2$ block determinant:
\begin{align}\label{rw2-tau}
&\tau_{n,k}=\det\left(\begin{matrix}
\tau_{n,k}^{[1,1]}&\tau_{n,k}^{[1,2]}\\
\tau_{n,k}^{[2,1]}&\tau_{n,k}^{[2,2]}
\end{matrix}\right),\\
&\tau_{n,k}^{[\overline{\alpha},\overline{\beta}]}=\left(m_{2s-1,2j-1}^{(n,k,\overline{\alpha},\overline{\beta})}\right)_{1\leq s\leq N_{{\overline{\alpha}}},1\leq j\leq N_{\overline{\beta}}},
\end{align}
for $\overline{\alpha},\overline{\beta}=1,2$, and the matrix elements ${m}_{s,j}^{(n,k,\overline{\alpha},\overline{\beta})}$ are given either
 (a) in differential operator form:
\begin{equation}\label{rw2-m}
\begin{aligned}
 {m}_{s,j}^{(n,k,\overline{\alpha},\overline{\beta})}=&\left.\frac{[\mathcal{T}(p)\partial_{p}]^{s}}{s!}\frac{[\widehat{\mathcal{T}}(\overline{p}^*)\partial_{\overline{p}^*}]^{j}}{j!}
m^{(n,k,\overline{\alpha},\overline{\beta})}\right|_{p=p_0^{(\overline{\alpha})},\overline{p}=p_0^{(\overline{\beta})}},\\
m^{(n,k,\overline{\alpha},\overline{\beta})}=&\frac{1}{{p}+\overline{p}^*}(-\frac{{p}-ik_1}{\overline{p}^{*}+ik_1})^n(-\frac{{p}-ik_2}{\overline{p}^{*}+ik_2})^ke^{\xi_{\overline{\alpha}}+\overline{\xi}_{\overline{\beta}}^*}.
\end{aligned}
\end{equation}
or (b) through Schur polynomials:
\begin{equation}\label{rw2-cm}
\begin{aligned}
 \widehat{{m}}_{s,j}^{(n,k,\overline{\alpha},\overline{\beta})}=&\sum\limits_{\nu=0}^{\mathrm{min}(i,j)}(\frac{1}{p_0^{(\overline{\alpha}) }+p_0^{(\overline{\beta})}})[\frac{|\lambda_{1,\alpha}|^2}{(p_0^{(\overline{\alpha})}+p_0^{(\overline{\beta})})^2}]^{\nu}\mathbf{S}_{i-\nu}({ \mathbf{x}^{+}_{\overline{\alpha},\overline{\beta}}}(n,k)+\nu{\bf s_{\overline{\alpha},\overline{\beta}}})
 \mathbf{S}_{j-\nu}({ \mathbf{x}^{-}_{\overline{\alpha},\overline{\beta}}}(n,k)+\nu{\bf s_{\overline{\beta},\overline{\alpha}}^*}),
\end{aligned}
\end{equation}
\noindent Here $N_1,N_2$ are arbitrary positive integers.
 The auxiliary functions in Eq. \eqref{rw2-m} are defined by
\begin{equation}\label{rw2-TQ}
\begin{aligned}
\mathcal{T}(p)=&\sqrt{\frac{\mathcal{Q}^2(p)-\mathcal{Q}^2(p_0^{(\overline{\alpha})})}{\mathcal{Q}^{'2}(p)}},
\widehat{\mathcal{T}}(\overline{p})=\sqrt{\frac{\mathcal{\widehat{Q}}^2(\overline{p})-\mathcal{\widehat{Q}}^2({p}_0^{(\overline{\beta})*})}{\mathcal{\widehat{Q}}^{'2}(\overline{p})}},\\
\xi_{\overline{\alpha}}=&px-\mathrm{i}p^2t+\sum\limits_{r=1}^{\infty}\widehat{a}_{r,\alpha}\ln^r\mathcal{W}^{(\overline{\alpha})}(p),
\overline{\xi}_{\overline{\beta}}=\overline{p}x-\mathrm{i}\overline{p}^2t+\sum\limits_{r=1}^{\infty}\widehat{a}^*_{r,\beta}\ln^r \mathcal{ \widehat{W}^{(\overline{\beta})}}(\overline{p}), \\
\mathcal{W}^{(\overline{\alpha})}(p)=&\frac{\mathcal{Q}(p)\pm\sqrt{\mathcal{Q}^2(p)-\mathcal{Q}(p^{(\overline{\alpha})}_0)}}{\mathcal{Q}(p^{(\overline{\alpha})}_0)},
\mathcal{\widehat{W}}^{(\overline{\beta})}(\overline{p})=\frac{\mathcal{\widehat{Q}}(\overline{p})\pm\sqrt{\mathcal{\widehat{Q}}^2(\overline{p})-\mathcal{\widehat{Q}}({p}^{(\overline{\beta})}_0)}}{\mathcal{\widehat{Q}}(p_0^{(\overline{\beta})*})}
,\\
\mathcal{\widehat{Q}}(\overline{p})=&\frac{\delta_1\rho_1^2}{\overline{p}+ik_1}+\frac{\delta_2\rho_2^2}{\overline{p}+ik_2}-i\overline{p}^2,
\end{aligned}
\end{equation}
where $\mathcal{Q}$ is given by Eq. \eqref{root-Q}.
 The vectors $\mathbf{x}_{\overline{\alpha},\overline{\beta}}^{\pm}(n,k)=(x_{1,\overline{\alpha},\overline{\beta}}^{\pm},x_{2,\overline{\alpha},\overline{\beta}}^{\pm},\cdots)$ in Eq. \eqref{rw2-cm} are defined as:
\begin{equation}\label{rw2-x}
\begin{aligned}
x_{r,\overline{\alpha},\overline{\beta}}^{+}(n,k)=&\lambda_{r,\overline{\alpha}}x+{i}\beta_{r,\overline{\alpha}}t+n\theta_{r,\overline{\alpha}}^{(1)}+k\theta_{r,\overline{\alpha}}^{(2)}-b_{r,\overline{\alpha},\overline{\beta}}+{a}_{r,\overline{\alpha}},  \\
x_{r,\overline{\alpha},\overline{\beta}}^{-}(n,k)=&\lambda_{r,\overline{\beta}}^*x-{i}\beta_{r,\overline{\beta}}^*t-n\theta_{r,\overline{\beta}}^{(1)}-k\theta_{r,\overline{\beta}}^{(2)}-b_{r,\overline{\beta},\overline{\alpha}}^*+{a}_{r,\overline{\beta}}^*,
\end{aligned}
\end{equation}
where $\lambda_{r,\overline{\alpha}},\beta_{r,\overline{\alpha}},\theta_{r,\overline{\alpha}}^{(1)},\theta_{r,\overline{\alpha}}^{(2)}$, and $b_{r,\overline{\alpha},\overline{\beta}}$ are coefficients from the expansions
\begin{equation}\label{rw2-abl}
\begin{aligned}
p(\kappa)-p_0^{(\overline{\alpha})}=&\sum\limits_{r=1}^{\infty}\lambda_{r,\overline{\alpha}}\kappa^{r},\\
p^2(\kappa)-p^{(\overline{\alpha})2}_0=&\sum\limits_{r=1}^{\infty}\beta_{r,\overline{\alpha}}\kappa^{r},\\
\ln\frac{p(\kappa)-ik_1}{p_0^{(\overline{\alpha})}-ik_1}=&\sum\limits_{r=1}^{\infty}\theta_{r,\overline{\alpha}}^{(1)}\kappa^{r},\\
\ln\frac{p(\kappa)-ik_2}{p_0^{(\overline{\alpha})}-ik_2}=&\sum\limits_{r=1}^{\infty}\theta_{r,\overline{\alpha}}^{(2)}\kappa^{r},\\
\ln\left[\frac{p^{(\overline{\alpha})}(\kappa)+p_0^{(\overline{\beta})}}{p_{0}^{(\overline{\alpha})}+p_0^{(\overline{\beta})}}\right]=&\sum\limits_{r=1}^{\infty}b_{r,\overline{\alpha},\overline{\beta}}\kappa^{r},\\
\mathcal{Q}[p^{(\overline{\alpha})}(\kappa)]=&\mathcal{Q}(p_0^{(\overline{\alpha})})\cosh(\kappa),
\end{aligned}
\end{equation}
and $\mathcal{Q}$ are defined by Eq. \eqref{rw1-TQ}.

\end{theorem}

\subsubsection{The dynamics of the mixed bounded $(N_{1},N_{2})$-th order
rogue waves}

This family of RW solutions is a mixture of a bounded $N_1$th-order RW and
another different bounded $N_2$th-order RW, thus it comprises of $\frac{%
N_1(N_1+1)+N_2(N_2+1)}{2}$ fundamental RWs. To exhibit the dynamics of the
mixed bounded RWs, here we take the two non-imaginary simple roots $%
p_0^{(\alpha)}$ ($\alpha=1,2$) of Eq. \eqref{root} and the parameters $%
\delta_{\alpha},k_{\alpha},\rho_{\alpha}$ as
\begin{equation}  \label{rw2-pra}
\begin{aligned}
&p_0^{(1)}=\frac{1}{2}+\frac{1}{3}i,p_0^{(2)}\approx0.5466-1.11i,
\delta_1=1,\delta_2=1,\\
&k_1=1,k_2=-1,\rho_1=\frac{25\sqrt{4002}}{1656},\rho_2=\frac{73%
\sqrt{46}}{552},h=0.
 \end{aligned}
\end{equation}

We first consider the simplest RWs in this solution family, which are
composed of two different fundamental RWs. For this purpose, we take $%
N_1=N_2=1$ in Theorem \ref{theorem-RW-2}. Then, the corresponding mixed
bounded RW solution is
\begin{equation}  \label{RW-2-1}
\begin{aligned} A=\rho_1e^{i\left(k_1x+(\gamma+k_1^2)t\right)}\frac{g}{f},
B=\rho_2e^{i\left(k_2x+(\gamma+k_2^2)t\right)}\frac{h}{f},L=\gamma-2( {\rm
log} f)_{xx}, \end{aligned}
\end{equation}
where
\begin{equation}  \label{RW-2-1fg}
\begin{aligned} f=&\begin{vmatrix}
\widehat{m}_{1,1}^{(0,0,1,1)}&\widehat{m}_{1,1}^{(0,0,1,2)}\\
\widehat{m}_{1,1}^{(0,0,2,1)}&\widehat{m}_{1,1}^{(0,0,2,2)} \end{vmatrix},
\quad g=&\begin{vmatrix}
\widehat{m}_{1,1}^{(1,0,1,1)}&\widehat{m}_{1,1}^{(1,0,1,2)}\\
\widehat{m}_{1,1}^{(1,0,2,1)}&\widehat{m}_{1,1}^{(1,0,2,2)}
\end{vmatrix},\quad h=&\begin{vmatrix}
\widehat{m}_{1,1}^{(0,1,1,1)}&\widehat{m}_{1,1}^{(0,1,1,2)}\\
\widehat{m}_{1,1}^{(0,1,2,1)}&\widehat{m}_{1,1}^{(0,1,2,2)} \end{vmatrix},
\end{aligned}
\end{equation}
and
\begin{equation}
\begin{aligned}
\widehat{m}_{1,1}^{(n,k,\overline{\alpha},\overline{\beta})}=&\frac{1}{p_{0}^{(\overline{\alpha})}+p_{0}^{(\overline{\beta})*}}
\left[x_{1,\overline{\alpha},\overline{\beta}}^{+}(n,k)x_{1,\overline{\alpha},\overline{\beta}}^{-}(n,k)+\frac{\lambda_{1,\overline{\alpha}}\lambda_{1,\overline{\beta}}^*}{(p_{0}^{(\overline{\alpha})}+p_{0}^{(\overline{\beta})*})^2}\right],\\
x_{\overline{\alpha},\overline{\beta}}^{+}(n,k)=&\alpha_{1,\overline{\alpha}}x+\mathrm{i}\beta_{1,\overline{\alpha}}t+(n+\frac{1}{2})\lambda_{1,\overline{\alpha}}+n_1\theta_{1,\overline{\alpha}}^{(1)}+n_2\theta_{1,\overline{\alpha}}^{(2)}-b_{1,\overline{\alpha},\overline{\beta}}+a_{1,\overline{\alpha}},  \\
x_{\overline{\alpha},\overline{\beta}}^{-}(n,k)=&\alpha_{1,\overline{\beta}}^*x-\mathrm{i}\beta_{1,\overline{\beta}}^*t-(n+\frac{1}{2})\lambda_{1,\overline{\beta}}^*-n_1\theta_{1,\overline{\beta}}^{(1)*}-n_2\theta_{1,\overline{\beta}}^{(2)*}-b_{1,\overline{\beta},\overline{\alpha}}^*+a_{1,\overline{\beta}}^*,
\end{aligned}
\end{equation}
for $\overline{\alpha},\overline{\beta}=1,2$. Here we have taken $%
a_0^{[1]}=1 $ and $a_0^{[2]}=1$ for simplicity. The degree of polynomials in
functions $f,g$, and $h$ is four in both $x$ and $t$.

\begin{figure}[!htb]
\centering
\subfigure{\includegraphics[height=3cm,width=14cm]{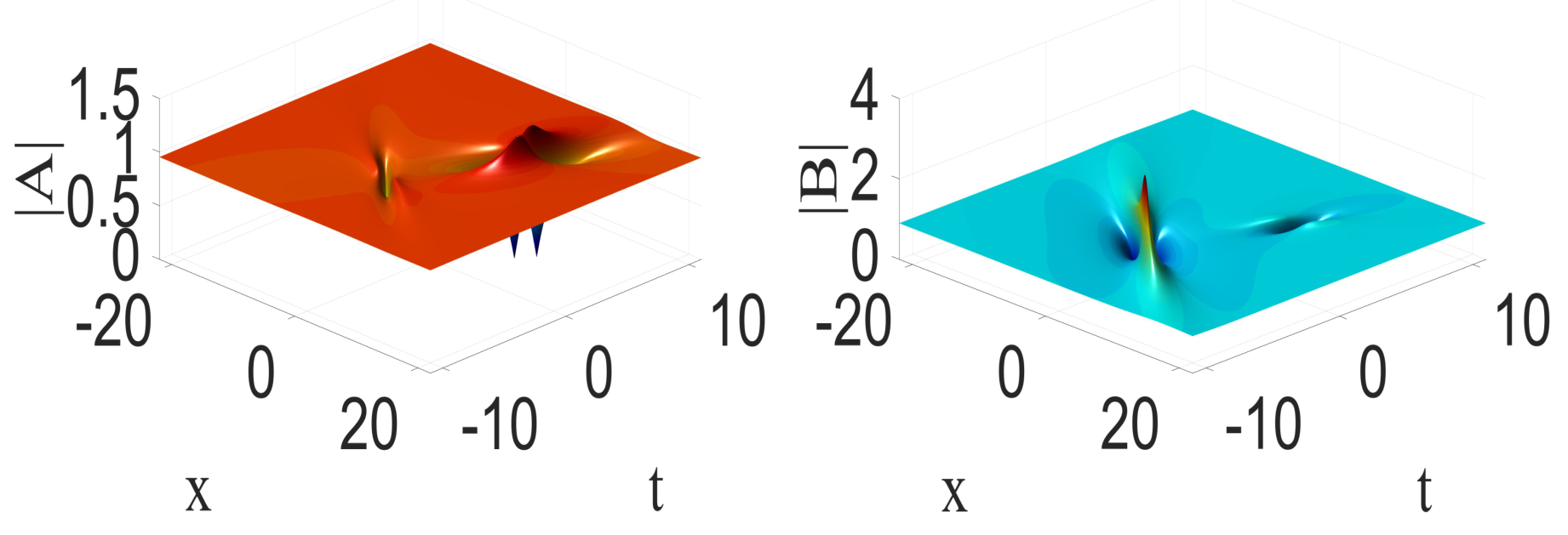}}
\caption{(Colour online) The mixed RW solution \eqref{RW-2-1} comprising of
two first-order RW with parameters given by Eq. \eqref{rw2-pra} and $%
N_1=1,N_2=1,a_{1,1}=0,a_{1,2}=10$.~}
\label{fig6}
\end{figure}

According to the classifications of the first-order RW solution \eqref{1RW-1-Final}
discussed previously, since $\frac{1}{3}%
p_{0R}^{(1)2}<(p_{0I}^{(1)}-k_1)^2<3p_{0R}^{(1)2},(p_{0I}^{(1)}-k_1)^2>{3}%
p_{0R}^{(2)2}$ and $(p_{0I}^{(2)}-k_1)^2<\frac{1}{3}%
p_{0R}^{(1)2},(p_{0I}^{(2)}-k_2)^2>{3}p_{0R}^{(2)2}$, thus the $A$ component
contains a dark RW and a four--petaled RW, while the $B$ component comprises
a bright RW and a dark RW. Fig. \ref{fig6} shows this mixed RW solution. It
is seen that the component $A $ features a four--petaled RW of larger shape
coexisting with a dark RW of smaller shape, while the component $B$
displays a dark RW of smaller shape mixing with a bright RW of bigger shape.

\begin{figure}[!htb]
\centering
\subfigure{\includegraphics[height=5cm,width=14cm]{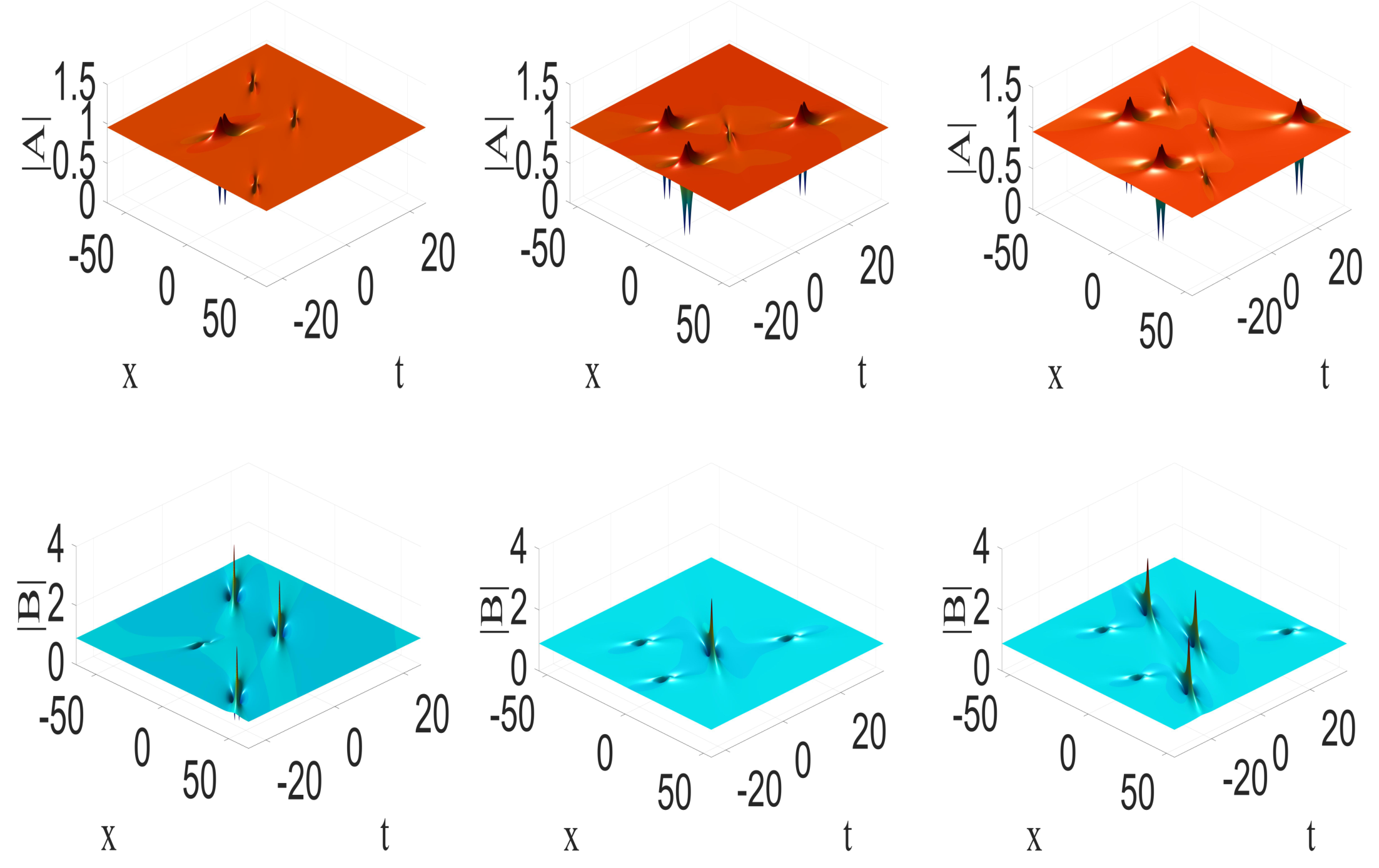}}
\caption{(Colour online) The leftmost side panels: the mixed RW solution
\eqref{RW-2} comprising of a first-order RW and a bounded
second-order RW with parameters given by Eq. \eqref{rw2-pra} and $%
N_1=1,N_2=2,a_{1,1}=0,a_{1,2}=0,a_{2,2}=0,a_{3,2}=400$. The middle panels:
the mixed RW solution \eqref{RW-2} comprising of a first-order RW
and a second-order bounded  RW with parameters given by Eq. \eqref{rw2-pra}
and $N_1=2,N_2=1,a_{1,1}=0,a_{2,1}=0,a_{3,1}=200i,a_{1,2}=0$. The rightmost
panels: the mixed RW solution \eqref{RW-2} consisting of two second-order bounded  RWs with
with parameters \eqref{rw2-pra} and $
N_1=2,N_2=2,a_{1,1}=0,a_{2,1}=0,a_{3,1}=500i,a_{1,2}=0,a_{2,2}=0,a_{3,2}=200.
$~}
\label{fig7}
\end{figure}

For larger $N_1$, or $N_2$, or both, the higher-order mixed bounded RWs can
be generated, which are composed of more fundamental RWs. For instance, by
taking $N_1=1,N_2=2$ or $N_1=2,N_2=1$ in Theorem \ref{theorem-RW-2}, the
corresponding solutions consist of a first-order RW and a second-order
bounded RW (i.e., four first-order RWs), and the degree of the corresponding
functions $f,g$, and $h$ are eight in both $x$ and $t$. The leftmost and
middle panels of Fig. \ref{fig7} display the mixed RWs with $N_1=1,N_2=2$
and $N_1=2,N_2=1$, respectively. It is seen that the component $B$ is
composed of three bounded fundamental RWs of bright type and a fundamental
dark RW when $N_1=1,N_2=2$ (see the leftmost panels of Fig. \ref{fig7}),
while it features three bounded fundamental RWs of dark type and a
fundamental bright RW when $N_1=2,N_2=1$ (see the middle panels of Fig. \ref%
{fig7} ). With $N_1=2,N_2=2$, the corresponding RW solutions %
\eqref{RW-2} are comprised of two bounded second-order RWs, which
are displayed in the rightmost panels of Fig. \ref{fig7}. It is seen that
there are three bounded fundamental bright RWs and three bounded fundamental
dark RWs in the $A$ component, and three bounded fundamental RWs of dark
type mixing with three bounded RWs of four--petaled type in the $A$
component.

It is noted that such mixed RWs were studied for the two component NLS
equations \cite{2NLS-1,2NLS-2,2NLS-3,2NLS-4,2NLS-5,2NLS-6}, but they have
not been reported for the 2-LSRI model before, to the best of our knowledge.
Additionally, in Theorem \ref{theorem-RW-2}, the parameters $N_1,N_2$ cannot
be zero, thus the $2\times2$ block determinant in Theorem \ref{theorem-RW-2}
cannot degenerate into a single block determinant. Therefore, the
corresponding mixed bounded RWs cannot reduce to pure bounded higher-order
RWs, i.e., they are always in mixed bounded states.

\subsection{The degradable bounded {$(\widehat{N}_{1},\widehat{N}_{2})$-th
order} rogue wave solutions and their dynamics}

\label{rw1-3}

\subsubsection{The degradable bounded {$(\widehat{N}_{1},\widehat{N}_{2})$%
-th order} rogue wave solutions}

If $\overline{p}_{0}$ is a non-imaginary double root of Eq. \eqref{root}, a family of the
degradable bounded RW solutions to the 2-LSRI model \eqref{2-LSRI} can be
constructed, which are given by the following Theorem.

\begin{theorem}\label{theorem-RW-3}
The 2-LSRI model \eqref{2-LSRI} has the following degradable bounded $(\widehat{N}_1,\widehat{N}_2)$th-order RW solutions
\begin{equation} \label{RW-3}
\begin{aligned}
A=\rho_1e^{i\left(k_1x+(\gamma+k_1^2)t\right)}\frac{g}{f}, B=\rho_2e^{i\left(k_2x+(\gamma+k_2^2)t\right)}\frac{h}{f},L=\gamma-2( {\rm log} f)_{xx},
\end{aligned}
\end{equation}
 where
\begin{equation} \label{rw3-fg}
\begin{aligned}
f=\tau_{0,0},g=\tau_{1,0},h=\tau_{0,1}
\end{aligned}
\end{equation}
and  $\tau_{n,k}$ is defined as the following $2\times2$ block determinant:
\begin{align}\label{rw3-tau-1}
&\tau_{n,k}=\det\left(\begin{matrix}
\tau_{n,k}^{[1,1]}&\tau_{n,k}^{[1,2]}\\
\tau_{n,k}^{[2,1]}&\tau_{n,k}^{[2,2]}
\end{matrix}\right),\\ \label{rw3-tau-2}
&\tau_{n,k}^{[\overline{\alpha},\overline{\beta}]}=\left(m_{3s-\overline{\alpha},3j-\overline{\beta}}^{[n,k,\overline{\alpha},\overline{\beta}]}\right)_{1\leq s\leq \widehat{N}_{\overline{\alpha}},1\leq j\leq \widehat{N}_{\overline{\beta}}},
\end{align}
for $\overline{\alpha},\overline{\beta}=1,2$,
and the matrix elements ${m}_{s,j}^{(n,k,\overline{\alpha},\overline{\beta})}$ are given either
(a) in differential operator form:
\begin{equation}\label{rw3-m}
\begin{aligned}
 {m}_{s,j}^{(n,k,\overline{\alpha},\overline{\beta})}=&\left.\sum\limits_{\mu=1}^{s}\sum\limits_{l=1}^{j}\frac{[\mathcal{T}(p)\partial_{p}]^{s}}{s!}\frac{[ {\mathcal{T}}^*( {p})\partial_{ {p}^*}]^{j}}{j!}m^{(n,k,\overline{\alpha},\overline{\beta})}\right|_{p=\overline{p}_0},\\
m^{(n,k,\overline{\alpha},\overline{\beta})}=&\frac{1}{{p}+{p}^*}(-\frac{{p}-ik_1}{{p}^{*}+ik_1})^n(-\frac{{p}-ik_2}{{p}^{*}+ik_2})^ke^{\xi_{\overline{\alpha}}+{\xi}^*_{\overline{\beta}}}.
\end{aligned}
\end{equation}
or (b) through Schur polynomials:
\begin{equation}\label{rw3-cm}
\begin{aligned}
 \widehat{{m}}_{s,j}^{(n,k,\overline{\alpha},\overline{\beta})}=&\sum\limits_{\nu=0}^{\mathrm{min}(i,j)}[\frac{|\lambda_1|^2}{(p_0+p_0^*)^2}]^{\nu}\mathbf{S}_{i-\nu}({ \mathbf{x}}_{\overline{\alpha}}^{+}(n,k)+\nu{\bf s})
 \mathbf{S}_{j-\nu}({ \mathbf{x}}_{\overline{\beta}}^{-}(n,k)+\nu{\bf s}^*).
\end{aligned}
\end{equation}
Here $\widehat{N}_1, \widehat{N}_2$ are arbitrary non-negative integers.
 The auxiliary functions $\mathcal{T}(p)$ in Eq. \eqref{rw3-m} is given through $\mathcal{Q}(p)$
\begin{equation} \label{rwT-3}
\begin{aligned}
\left(\mathcal{T}(p)\partial_{p}\right)^3\mathcal{Q}(p)=\mathcal{Q}(p),
\end{aligned}
\end{equation}
and $\xi_{\overline{\alpha}}$ is defined by
\begin{equation}\label{rw3-xi}
\begin{aligned}
\xi_{\overline{\alpha}}=&px-ip^2t+\sum\limits_{r=1}^{\infty}a_{r,\overline{\alpha}}\ln^{r}\mathcal{W}(p).
\end{aligned}
\end{equation}
The vectors $\mathbf{x}_{\overline{\alpha}}^{\pm}(n,k)=(x_{1,\overline{\alpha}}^{\pm},x_{2,\overline{\alpha}}^{\pm},\cdots)$ in Eq. \eqref{rw3-cm} are defined as:
\begin{equation}\label{rw3-x}
\begin{aligned}
x_{r,\overline{\alpha}}^{+}(n,k)=&\lambda_{r,\overline{\alpha}}x+{i}\beta_{r,\overline{\alpha}}t+n\theta_{r,\overline{\alpha}}^{(1)}+k\theta_{r,\overline{\alpha}}^{(2)}+{a}_{r,\overline{\alpha}},  \\
x_{r,\overline{\alpha}}^{-}(n,k)=&\lambda_{r,\overline{\alpha}}^*x-{i}\beta_{r,\overline{\alpha}}^*t-n\theta_{r,\overline{\alpha}}^{(1)}-k\theta_{r,\overline{\alpha}}^{(2)}+{a}_{r,\overline{\alpha}}^*,
\end{aligned}
\end{equation}
where $\alpha_r,\beta_r,\theta_r^{(1)}$, and $\theta_r^{(2)}$ are defined in Eq. \eqref{rw1-abl} with $p_0$ replaced by $\overline{p}_0$, $\mathbf{s}=(s_1,s_2,s_3,\cdots)$ is given by Eq. \eqref{rw1-s} with $p_0$ replaced by $\overline{p}_0$, and the function $p(\kappa)$ appearing in Eqs. \eqref{rw1-abl},\eqref{rw1-s} are defined by the following equation:
\begin{equation}\label{rw3-pq}
\begin{aligned}
\mathcal{Q}(p)=\frac{\mathcal{Q}(p_0)}{3}\left[e^{\kappa}+2e^{-\frac{\kappa}{2}}\cos(\frac{\sqrt{3}}{2}\kappa)\right],
\end{aligned}
\end{equation}
where $\mathcal{Q}$ is given by Eq. \eqref{root-Q}.

\end{theorem}

\begin{remark}
By using the same method developed in Ref. \cite{orw-2}, one can get that the polynomial degree of the tau functions $\tau_{n,k}$ in Theorem \ref{theorem-RW-1}--\ref{theorem-RW-3} are $N(N+1)$,  $N_1(N_1+1)+N_2(N_2+1)$ and  $2[\widehat{N}_1^2+\widehat{N}_2^2-\widehat{N}_1(\widehat{N}_2-1)]$
 in both $x$ and $t$ variables, respectively, where $N,N_1,N_2$ are positive integers and $\widehat{N}_1,\widehat{N}_2$ are non-negative integers.
\end{remark}

\subsubsection{Dynamics of the degradable bounded rogue waves}

Similar to the mixed bounded RWs in Theorem \ref{theorem-RW-2}, the RW
solutions in the above Theorem are also given by $2\times2$ block
determinants. However, the $2\times2$ block determinants in this Theorem can
degenerate into single block determinants if one takes $\widehat{N}_1=0$ or $%
\widehat{N}_2=0$. That is the reason we call this family of solutions degradable RW
solutions. In what follows, we will reveal the dynamics of the degenerate
single-block RW solutions and the non-degenerate $2\times2 $ block
solutions, respectively. For this purpose, we take the following parameters
in the solutions \eqref{RW-3} of Theorem \ref{theorem-RW-3}
\begin{equation}  \label{pa-de}
\begin{aligned}
&\overline{p}_0=\frac{1}{4}(\sqrt{15}+3i),\rho_1=\frac{2\sqrt{6}}{3},
\rho_2=\frac{5\sqrt{3}}{3},\delta_1=-1,\delta_2=-1,k_1=1,k_2=-\frac{1}{2}.
\end{aligned}
\end{equation}
With this set of parameters, upon the classifications of the first-order RW solutions %
\eqref{1RW-1-Final} discussed previously, the solutions in the components $A$ and $%
B$ are dark RWs and bright RWs, respectively.

We first consider the degenerate single-block RW solutions (i.e., $\widehat{N%
}_1=0$ or $\widehat{N}_2=0$). In the degenerate case of $\widehat{N}_1=0,%
\widehat{N}_2\neq0$ and the degenerate case of $\widehat{N}_1\neq0,\widehat{%
N}_2=0$, the dynamical features of the corresponding RWs are diverse. In
order to more clearly show the difference between the two degenerate RWs,
we next exhibit them.

\begin{figure}[!htb]
\centering
\subfigure{\includegraphics[height=5cm,width=14cm]{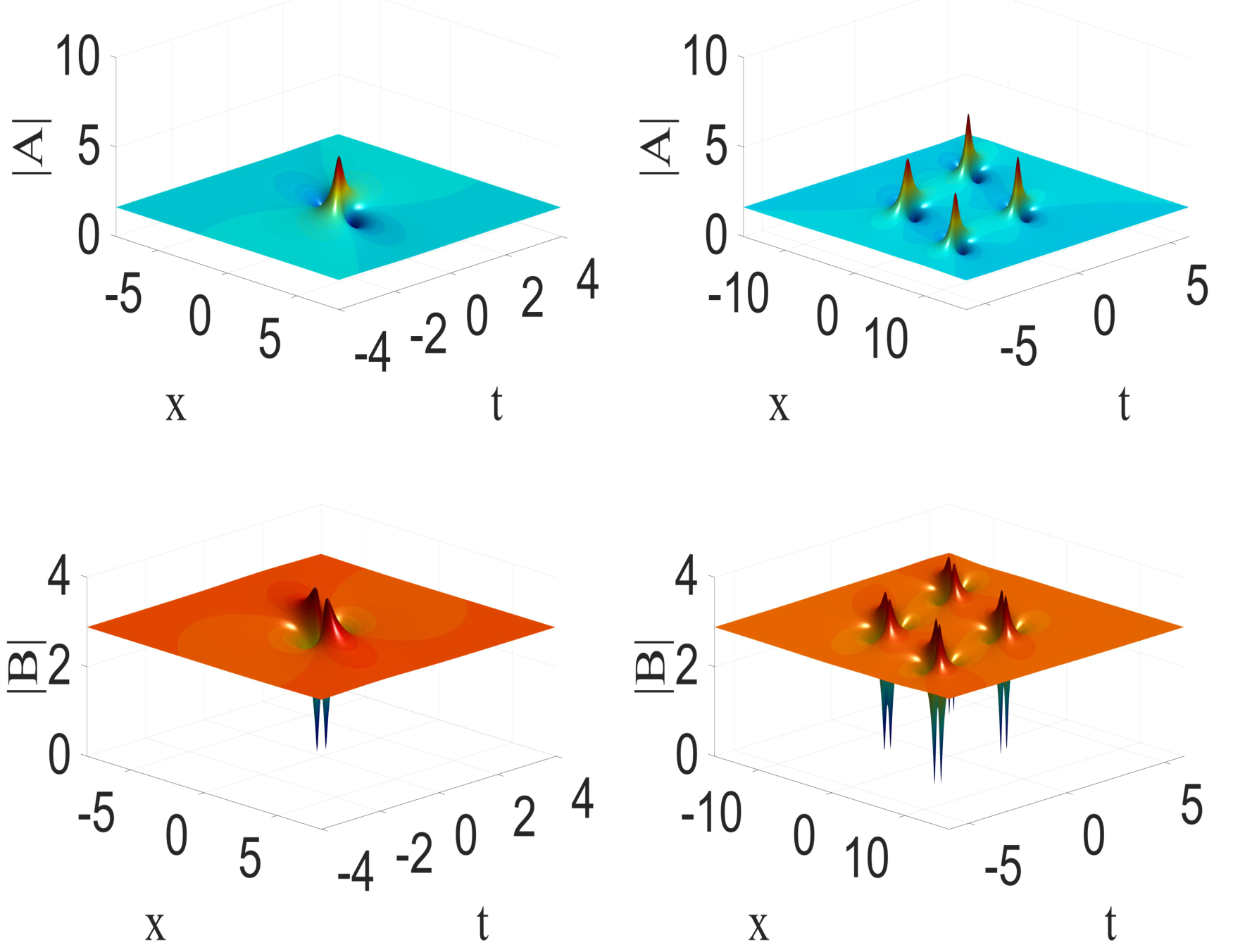}}
\caption{(Colour online) The degenerate RW solutions \eqref{RW-3} with $%
\widehat{N}_1=0$ and parameters given by Eq. \eqref{pa-de}. The left panels:
$\widehat{N}_2=1$ and $a_{1,2}=0$, the corresponding solutions only comprise
a single fundamental RW. The {\ right} panels: $\widehat{N}_2=2$ and $%
a_{1,2}=0,a_{2,2}=0,a_{3,2}=0,a_{4,2}=100$, the corresponding solutions
consist of four fundamental RWs. ~}
\label{fig8}
\end{figure}

When $\widehat{N}_1=0,\widehat{N}_2\neq0$, the tau solution in Eq. %
\eqref{rw3-tau-1} has the following form:
\begin{equation}  \label{N01-tau}
\begin{aligned}
\tau_{n,k}=\det\left(\tau_{n,k}^{[2,2]}\right)=\det\left(%
\widehat{m}_{3s-2,3j-2}^{(n,k,2,2)}\right)_{1\leq s,j\leq \widehat{N}_2},
\end{aligned}
\end{equation}
with $\widehat{m}_{s,j}^{(n,k,\alpha,\beta)}$ being given by Eq. %
\eqref{rw3-cm}. In this degenerate case, the parameter $\widehat{N}_2$
determines the order of the degenerate RWs, the corresponding solutions
consist of $\widehat{N}_2^2$ fundamental RWs. Fig. \ref{fig8} shows the
degenerate RWs for $\widehat{N}_2=1$ and $\widehat{N}_2=2$. It is seen that
there is only one fundamental RW in the degenerate solutions when $\widehat{N%
}_2=1$ (see the left panels). In this case, the corresponding degenerate RW
solution is equivalent to the fundamental RW solutions given by Eq. %
\eqref{1RW-1}. However, in the case of $\widehat{N}_2=2$, there are four
fundamental RWs (see the right panels). As discussed previously, the mixed
bounded RWs can also form a RW pattern consisting of four fundamental RWs
(see the leftmost side and middle panels of Fig. \ref{fig7}). However, the
four fundamental RWs in the mixed bounded solutions are composed of three
bounded RWs and a fundamental RW of different state (see the leftmost side
and middle panels of Fig. \ref{fig7}). In the degenerate solution, the four
fundamental RWs are bounded (see the right panels of Fig. \ref{fig8}).

\begin{figure}[!htb]
\centering
\subfigure{\includegraphics[height=5cm,width=14cm]{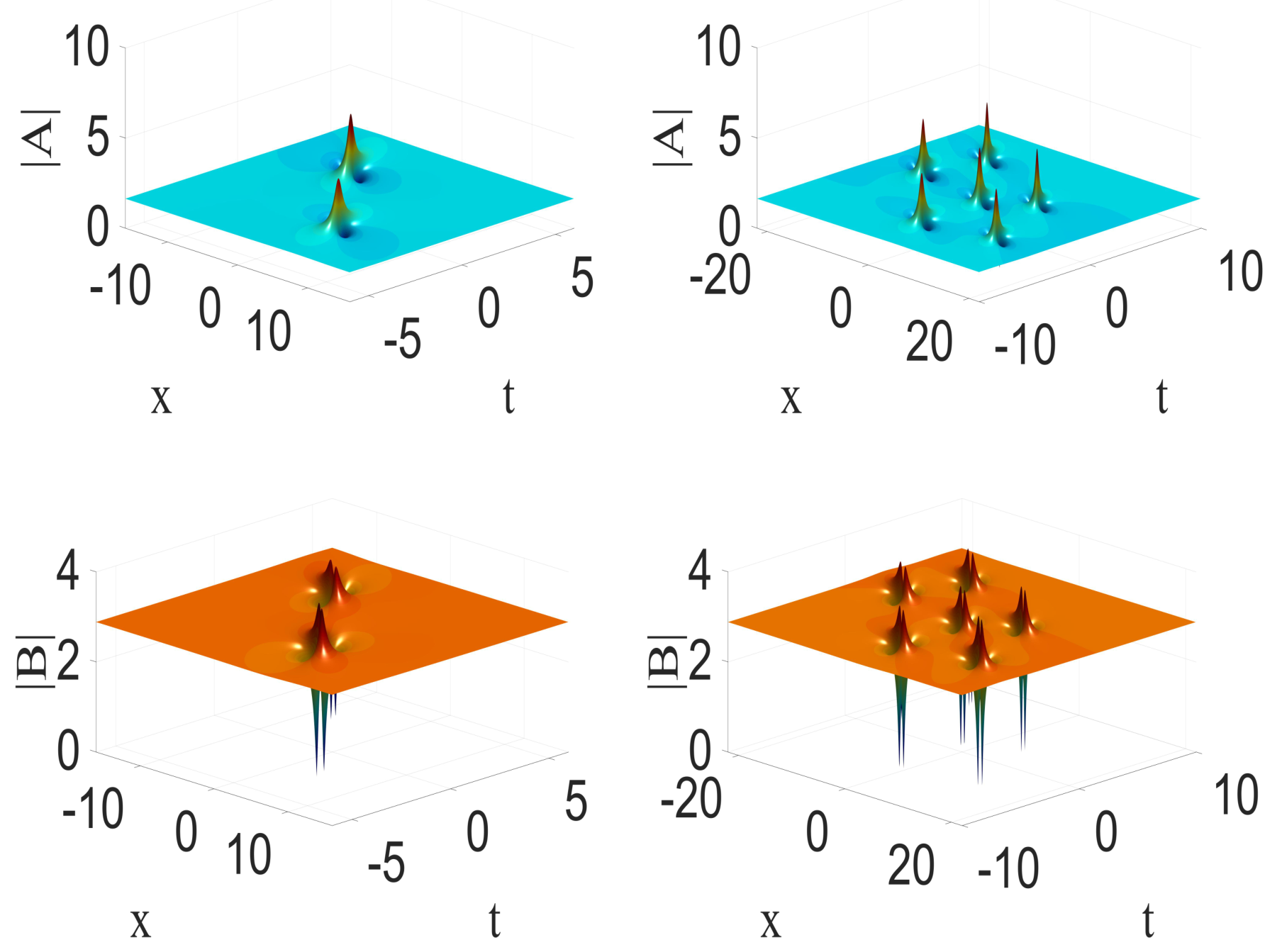}}
\caption{(Colour online) The degenerate RW solutions \eqref{RW-3} with $%
\widehat{N}_2=0$ and parameters given by Eq. \eqref{pa-de}. The left panels:
$\widehat{N}_1=1$ and $a_{1,1}=0,a_{2,1}=10$, the corresponding solutions
consist of two bounded fundamental RWs. The right panels: $\widehat{N}_1=2$
and $a_{1,1}=0,a_{2,1}=0,a_{3,1}=0,a_{4,1}=0,a_{5,1}=500$, the corresponding
solutions consist of six bounded fundamental RWs. ~ }
\label{fig9}
\end{figure}

When $\widehat{N}_1\neq0,\widehat{N}_2=0$, the tau solution in Eq. %
\eqref{rw3-tau-1} is written in the following form:
\begin{equation}  \label{N10-tau}
\begin{aligned}
\tau_{n,k}=\det\left(\tau_{n,k}^{[1,1]}\right)=\det\left(%
\widehat{m}_{3s-1,3j-1}^{(n,k,1,1)}\right)_{1\leq s,j\leq \widehat{N}_1},
\end{aligned}
\end{equation}
where $\widehat{m}_{s,j}^{(n,k,\alpha,\beta)}$ is given by Eq. \eqref{rw3-cm}%
. In this case, the parameter $\widehat{N}_1$ controls the order of the
corresponding degenerate RW, and there are $(\widehat{N}_1^2+\widehat{N}_1)$
fundamental RWs in degenerate RW solutions. The RW solutions \eqref{RW-3}
with $\widehat{N}_1=1$ and $\widehat{N}_1=2$ are displayed in Fig. \ref{fig9}%
. It is seen that there are two bounded fundamental RWs in the case $\widehat{N}_1=1$
(see the left panels of Fig. \ref{fig9}) while there are six bounded
fundamental RWs when $\widehat{N}_1=2$ (see the right panels of Fig. \ref{fig9}). The
solutions shown in Fig. \ref{fig6} are also composed of two fundamental RWs,
but they are in different states.

\begin{figure}[!htb]
\centering
\subfigure{\includegraphics[height=3cm,width=14cm]{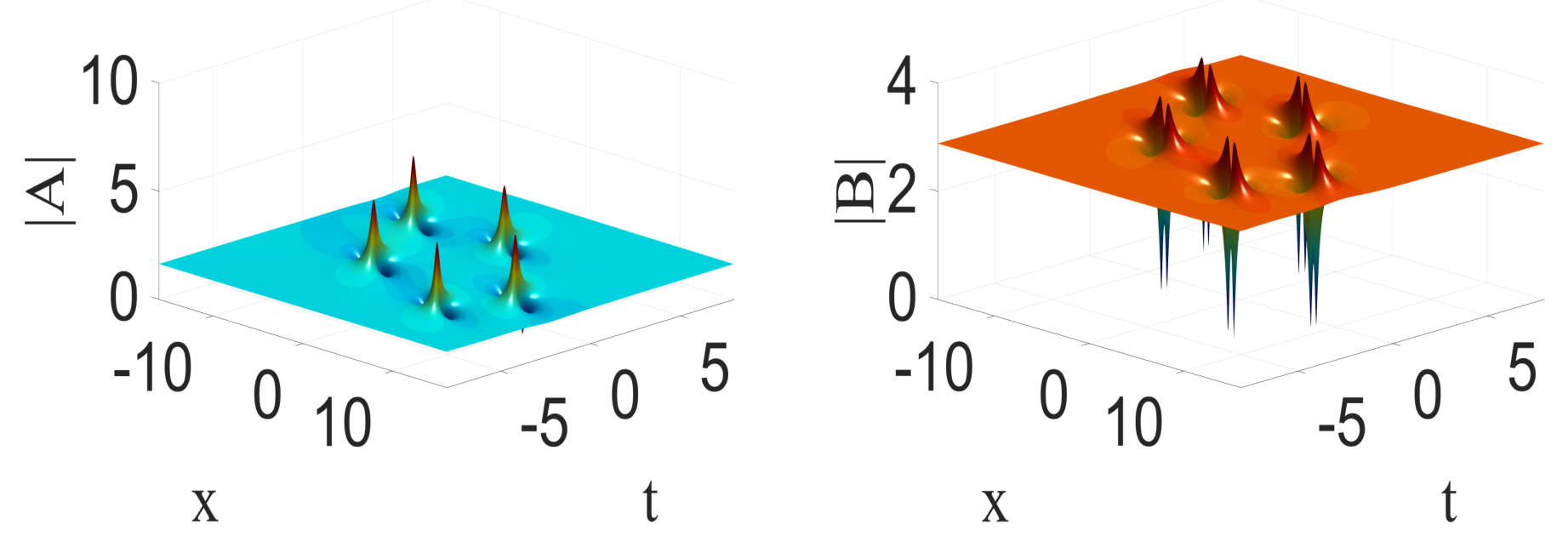}}
\caption{(Colour online) The non-degenerate RW solutions \eqref{RW-3} with $%
\widehat{N}_1=2,\widehat{N}_2=1$ and parameters \eqref{pa-de} and $%
a_{1,1}=0,a_{2,1}=0,a_{3,1}=0,a_{4,1}=0,a_{5,1}=500,a_{1,2}=0$, which
consist of five bounded fundamental RWs. ~}
\label{fig10}
\end{figure}

Then we consider the non-degenerate RWs solutions when $\widehat{N}_1%
\widehat{N}_2\neq0$, which are expressed by the $2\times2$ block determinants.
To illustrate the dynamics of these non-degenerate RWs, we consider the case
of $\widehat{N}_1=2,\widehat{N}_2=1$. In this case, the tau function is
explicitly expressed as
\begin{align}
\tau_{n,k}=&\det\left(%
\begin{matrix}
\tau_{n,k}^{[1,1]} & \tau_{n,k}^{[1,2]} \\
\tau_{n,k}^{[2,1]} & \tau_{n,k}^{[2,2]}%
\end{matrix}%
\right), \\
=&\det\left(%
\begin{matrix}
\widehat{m}_{2,2}^{(n,k,1,1)} & \widehat{m}_{2,5}^{(n,k,1,1)} & \widehat{m}%
_{2,1}^{(n,k,1,2)} \\
\widehat{m}_{5,2}^{(n,k,1,1)} & \widehat{m}_{5,5}^{(n,k,1,1)} & \widehat{m}%
_{5,1}^{(n,k,1,2)} \\
\widehat{m}_{1,2}^{(n,k,2,1)} & \widehat{m}_{1,5}^{(n,k,2,1)} & \widehat{m}%
_{1,1}^{(n,k,2,2)}%
\end{matrix}%
\right), \\
\end{align}
where $\widehat{m}_{s,j}^{(n,k,\alpha,\beta)}$ is given in Eq. \eqref{rw3-cm}%
. The degree of these tau functions is ten in both $x$ and $t$, thus the
corresponding non-degenerate RW should be composed of five fundamental RWs,
which is demonstrated in Fig. \ref{fig10}. Since the solutions in Theorems %
\ref{theorem-RW-1},\ref{theorem-RW-2} consist of $\frac{N(N+1)}{2}$ and $%
\frac{N_1(N_1+1)+N_2(N_2+1)}{2}$ fundamental RWs, respectively, which do not
contain the solutions comprising five fundamental RWs, thus the RW solution
displayed in Fig. \ref{fig10} is distinctive in contrast with the solutions
in Theorems \ref{theorem-RW-1},\ref{theorem-RW-2}.

\section{Derivation of higher-order breather and RW solutions}

\label{derivation} In this Section, we construct the general breather in
Theorem \ref{theorem1} and RW solutions in Theorems \ref{theorem-RW-1},\ref%
{theorem-RW-2},\ref{theorem-RW-3} to the 2-LSRI model \eqref{2-LSRI} via the
bilinear KP hierarchy reduction method \cite{Hirota,jimbo-1983,Kyoto}.

\subsection{Tau functions of the 2-LSRI model}

\label{derivation-breather} The 2-LSRI model \eqref{2-LSRI} is transformed
into the bilinear form,
\begin{equation}  \label{Bi-2-LSRI}
\begin{aligned} &\left(D_{x}^2+2ik_{1}D_x-iD_{t}\right)g\cdot f =0,\\
&\left(D_{x}^2+2ik_{2}D_x-iD_{t}\right)h\cdot f =0,\\
&\left(D_tD_x-2\delta_1\rho_1^2-2\delta_2\rho_2^2\right)f\cdot
f=-2(\delta_1\rho_1^2|g^2|+\delta_2\rho_2^2|h|^2), \end{aligned}
\end{equation}
through the variable transformation,
\begin{equation}  \label{Bi-Tr}
\begin{aligned} A=\rho_1e^{i\left(k_1x+(\gamma+k_1^2)t\right)}\frac{g}{f},
B=\rho_2e^{i\left(k_2x+(\gamma+k_2^2)t\right)}\frac{h}{f},L=\gamma-2( {\rm
log} f)_{xx}, \end{aligned}
\end{equation}
where $f$ is a real function, and $g$, $h$ are complex functions. Here the
operator $D$ is the Hirota's bilinear differential operator \cite{Hirota}
defined by
\begin{equation}
\begin{aligned} &P(D_{x},D_{y},D_{t}, )F(x,y,t\cdot\cdot\cdot)\cdot
G(x,y,t,\cdot\cdot\cdot)\\
=&P(\partial_{x}-\partial_{x^{'}},\partial_{y}-\partial_{y^{'}},
\partial_{t}-\partial_{t^{'}},\cdot\cdot\cdot)F(x,y,t,\cdot\cdot%
\cdot)G(x^{'},y^{'},t^{'},\cdot\cdot\cdot)|_{x^{'}=x,y^{'}=y,t^{'}=t},%
\nonumber \end{aligned}
\end{equation}
where $P$ is a polynomial of $D_{x}$,$D_{y}$,$D_{t},\cdot\cdot\cdot$.

Then we start with the general tau functions for the multi-component KP
hierarchy expressed in the forms of Gramian determinants. The following
bilinear equation in the multi-component KP hierarchy \cite{ohta-2011}
\begin{equation}  \label{KPeq}
\begin{aligned} &(D_{x_1}^{2}+2\alpha_1D_{x_1}-D_{x_2})\tau_{n+1,k} \cdot
\tau_{n,k} =0,\\ &(D_{x_1}^{2}+2\alpha_2D_{x_1}-D_{x_2})\tau_{n,k+1} \cdot
\tau_{n,k} =0,\\ &(D_{x_1}D_{r}-2)\tau_{n,k} \cdot \tau_{n,k}
=-2\tau_{n+1,k}\tau_{n-1,k},\\ &(D_{x_1}D_{s}-2)\tau_{n,k} \cdot \tau_{n,k}
=-2\tau_{n,k+1}\tau_{n,k-1},\\ \end{aligned}
\end{equation}
has the following Gramian determinant tau functions
\begin{equation}  \label{tau}
\begin{aligned} \tau_{n,k}=\det\limits_{1\leq s,j\leq N} (m_{s,j}^{(n,k)}).
\end{aligned}
\end{equation}
Here the matrix element $m_{ij}^{(n)}$ satisfies
\begin{equation}  \label{eq:m}
\begin{aligned}
\partial_{x_{1}}m_{s,j}^{(n,k)}&=\psi_{s}^{(n,k)}\phi_{j}^{(n,k)}, \\
\partial_{x_{2}}m_{s,j}^{(n,k)}&=\psi_{s}^{(n+1,k)}\phi_{j}^{(n,k)}+%
\psi_{s}^{(n,k)}\phi_{j}^{(n-1,k)},\partial_{x_{2}}m_{s,j}^{(n)}=%
\psi_{s}^{(n)}\phi_{j}^{(n,k+1)}+\psi_{s}^{(n,k)}\phi_{j}^{(n,k-1)}, \\
m_{s,j}^{(n+1,k)}&=m_{s,j}^{(n,k)}+\psi_{s}^{(n,k)}%
\phi_{j}^{(n+1,k)},m_{s,j}^{(n,k+1)}=m_{s,j}^{(n,k)}+\psi_{s}^{(n,k)}%
\phi_{j}^{(n,k+1)}, \\
\partial_{x_{2}}\psi_{s}^{{(n,k)}}&=\partial_{x_1}^2\psi_{s}^{(n,k)}, \\
\psi_{s}^{{(n+1,k)}}&=(\partial_{x_1}-\alpha_1)\psi_{s}^{(n,k)},
\psi_{s}^{{(n,k+1)}}=(\partial_{x_1}-\alpha_2)\psi_{s}^{(n,k)}, \\
\partial_{x_{2}}\phi_{j}^{{(n,k)}}&=-\partial_{x_1}^2\phi_{s}^{(n,k)}, \\
\phi_{s}^{{(n+1,k)}}&=-(\partial_{x_1}+\alpha_1)\phi_{s}^{(n,k)},
\phi_{s}^{{(n,k+1)}}=-(\partial_{x_1}+\alpha_2)\phi_{s}^{(n,k)},
\end{aligned}
\end{equation}
where $m_{s,j}^{(n,k)},\phi_s^{(n,k)}$, and $\phi_j^{(n,k)}$ are variables
of $x_1,x_2,r$, and $s$.

If one restrict the above tau functions satisfying the following
dimension-reduction condition:
\begin{equation}  \label{dimension}
\begin{aligned}
\mathcal{L}_0\tau_{n,k}=(\delta_1\rho_1^2\partial_s+\delta_2\rho_2^2%
\partial_s+i\partial_{x_2})\tau_{n,k}=C\tau_{n,k}, \end{aligned}
\end{equation}
where
\begin{equation}  \label{dimension-L}
\begin{aligned}
\mathcal{L}_0=(\delta_1\rho_1^2\partial_s+\delta_2\rho_2^2\partial_s+i%
\partial_{x_2}), \end{aligned}
\end{equation}
and $C$ is some constant, then the third and fourth bilinear equations in
Eq. \eqref{KPeq} would combine into the following bilinear equation:
\begin{equation}  \label{Bi-2-LSRIX}
\begin{aligned}
\left(D_{x_1}D_{x_2}-2\delta_1-2\delta_2\right)\tau_{n,k}\cdot
\tau_{n,k}=-2(\delta_1\tau_{n+1,k}\tau_{n-1,k}+\delta_2\tau_{n,k+1}%
\tau_{k-1}). \end{aligned}
\end{equation}
Based on this dimension-reduction condition, the derivatives with respect to
variables $r$ and $s$ are replaced by the derivative with respect to another
variables $x_1,x_2$. Then in the above tau functions, $s$ and $r$ are just
parameters that can be regarded as having any values, thus we take $s=r=0$
for convenience. Applying the change of independent variables in the tau
functions \eqref{tau}
\begin{equation}  \label{variable}
\begin{aligned} x_1=x,x_2=-it, \alpha_{1}=ik_1,\alpha_2=ik_2, \end{aligned}
\end{equation}
if the tau functions $\tau_{n,k}$ satisfy the complex conjugacy condition:
\begin{equation}  \label{com-con}
\begin{aligned} \tau_{n,k}^*(x,t)=\tau_{-n,-k}(x,t), \end{aligned}
\end{equation}
then the bilinear equation \eqref{Bi-2-LSRIX} and the first and second
bilinear equations in Eq. \eqref{KPeq} would become the bilinear equations %
\eqref{Bi-2-LSRI} of the 2-LSRI model for
\begin{equation}
f=\tau_{0,0},g=\tau_{1,0},g^*=\tau_{-1,0},h=\tau_{0,1},h^*=\tau_{0,-1}.
\end{equation}
In this way, we obtain the tau functions of the 2-LSRI model.

Choosing different forms of the matrix elements $m_{s,j}^{(n,k)}$, we can
construct the breather and RW solutions for the 2-LSRI model \eqref{2-LSRI},
the details of the derivations will be given in the following subsections.

\subsection{Derivation of breather solutions to the 2-LSRI model}

To construct the breather solutions of the 2-LSRI model \eqref{2-LSRI}, we
take the matrix elements $m_{s,j}^{(n,k)}$ of tau functions $\tau_{n,k}$
given by Eq. \eqref{tau} in the following form:
\begin{equation}
\begin{aligned}
m_{s,j}^{(n,k)}=&\sum\limits_{\alpha,\beta=1}^{2}\frac{1}{p_s^{[%
\alpha]}+q_j^{[\beta]}}(-\frac{p_s^{[\alpha]}-\alpha_1}{q_j^{[\beta]}+%
\alpha_1})^n (-\frac{p_s^{[\alpha]}-\alpha_2}{q_j^{[\beta]}+\alpha_2})^k
e^{\xi_s^{[\alpha]}+\eta_j^{[\beta]}}, \\
\psi_s^{(n,k)}=&\sum\limits_{\alpha=1}^2(p_s^{[\alpha]}-\alpha_1)^{n}(p_s^{[%
\alpha]}-\alpha_2)^{k}e^{\xi_s^{[\alpha]}}, \\
\phi_j^{(n,k)}=&\sum\limits_{\beta=1}^2(-q_j^{[\beta]}-%
\alpha_1)^{-n}(-q_j^{[\beta]}-\alpha_2)^{-k}e^{\eta_j^{[\beta]}},
\end{aligned}
\end{equation}
and
\begin{equation}
\begin{aligned}
\xi_s^{[\alpha]}=&\frac{1}{p_s^{[\alpha]}-\alpha_1}r+\frac{1}{p_s^{[%
\alpha]}-\alpha_2}s+p_s^{[\alpha]}x_1+p_s^{[\alpha]2}x_2+\overline{\xi}_s^{[%
\alpha]},\\
\eta_s^{[\beta]}=&-\frac{1}{q_j^{[\beta]}+\alpha_1}r-\frac{1}{q_j^{[\beta]}+%
\alpha_2}s+q_j^{[\beta]}x_1-q_j^{[\beta]2}x_2+\overline{\eta}_j^{[\beta]},
\end{aligned}
\end{equation}
where $p_{s}^{[\alpha]},q_j^{[\beta]},\overline{\xi}_s^{[\alpha]}$, and $%
\overline{\eta}_j^{[\beta]}$ are arbitrary complex parameters. Then the tau
function in Eq. \eqref{tau} can be rewritten in the following form:
\begin{equation}  \label{tau-1}
\begin{aligned} \tau_{n,k}=\Lambda\det\limits_{1\leq s,j\leq
N}\left(\overline{m}_{s,j}^{(n,k)}\right), \end{aligned}
\end{equation}
where $\Lambda=\prod\limits_{s=1}^{N}e^{\xi_s^{[2]}+\eta_j^{[2]}}$ and
\begin{equation}  \label{m-2}
\begin{aligned}
\overline{m}_{s,j}^{(n,k)}=&\frac{1}{p_s^{[1]}+q_j^{[1]}}(-\frac{p_s^{[1]}-%
\alpha_1}{q_j^{[1]}+\alpha_1})^n
(-\frac{p_s^{[1]}-\alpha_2}{q_j^{[1]}+\alpha_2})^k
e^{\xi_s^{[1]}-\xi_s^{[2]}+\eta_j^{[1]}-\eta_j^{[2]}}+%
\frac{1}{p_s^{[1]}+q_j^{[2]}}(-\frac{p_s^{[1]}-\alpha_1}{q_j^{[2]}+%
\alpha_1})^n\\ &\times(-\frac{p_s^{[1]}-\alpha_2}{q_j^{[2]}+\alpha_2})^k
e^{\xi_s^{[1]}-\xi_s^{[2]}}+\frac{1}{p_s^{[2]}+q_j^{[1]}}(-\frac{p_s^{[2]}-%
\alpha_1}{q_j^{[1]}+\alpha_1})^n(-\frac{p_s^{[2]}-\alpha_2}{q_j^{[1]}+%
\alpha_2})^k e^{\eta_j^{[1]}-\eta_j^{[2]}}+\\
&\frac{1}{p_s^{[2]}+q_j^{[2]}}(-\frac{p_s^{[2]}-\alpha_1}{q_j^{[2]}+%
\alpha_1})^n(-\frac{p_s^{[2]}-\alpha_2}{q_j^{[2]}+\alpha_2})^k. \end{aligned}
\end{equation}
If the parameters $p_s^{[\alpha]},q_j^{[\beta]}$ satisfy the constraints
\begin{equation}  \label{constraint-1}
\begin{aligned}
\frac{\delta_1\rho_1^2}{(p_s^{[1]}-\alpha_1)(p_s^{[2]}-\alpha_2)}+\frac{%
\delta_2\rho_2^2}{(p_s^{[1]}-\alpha_2)(p_s^{[2]}-%
\alpha_2)}-i(p_s^{[1]}+p_s^{[2]})=0, \\
\frac{\delta_1\rho_1^2}{(q_s^{[1]}+\alpha_1)(q_s^{[2]}+\alpha_1)}+\frac{%
\delta_2\rho_2^2}{(q_s^{[1]}+\alpha_2)(q_s^{[2]}+%
\alpha_2)}+i(q_s^{[1]}+q_s^{[2]})=0, \end{aligned}
\end{equation}
then from Eqs. \eqref{tau-1}, \eqref{m-2}, we get the tau function in the
form of Eq. \eqref{tau-1} satisfying the dimension-reduction condition %
\eqref{dimension}. By taking the variable transformations \eqref{variable}
into the tau functions \eqref{tau-1}, and setting the following complex
conjugacy conditions on the parameters:
\begin{equation}  \label{pq-1}
\begin{aligned} q_j^{[\beta]}=p_j^{[\beta]*},
\overline{\eta}_j^{[\beta]}=\overline{\xi}_s^{[\alpha]*}, \end{aligned}
\end{equation}
then we have
\begin{equation}
\begin{aligned}
\eta_j^{[1]}-\eta_j^{[2]}=\xi_s^{[1]*}-%
\xi_s^{[2]*},m_{s,j}^{(n,k)*}=m_{j,s}^{(-n,-k)}, \end{aligned}
\end{equation}
which further yields the tau function \eqref{tau-1} satisfying the complex
conjugacy condition \eqref{com-con}.

Thus, under the parameter constraint in Eq. \eqref{constraint-1} and the
variable transformations in Eq. \eqref{variable}, the tau functions %
\eqref{tau-1} reduce to the solutions of bilinear equations \eqref{Bi-2-LSRI}
of the 2-LSRI model \eqref{2-LSRI} for $f=\tau_{0,0},g=\tau_{1,0},g^*=%
\tau_{-1,0},h=\tau_{0,1},h^*=\tau_{0,-1}$. The multiplicative factor $%
\Lambda $ in Eq. \eqref{tau-1} can be removed in $\frac{\tau_{1,0}}{%
\tau_{0,0}}$ and $\frac{\tau_{0,1}}{\tau_{0,0}}$, and the variables $s$,$r$
are taken as zero, then we can obtain the breather solutions of the 2-LSRI
model \eqref{2-LSRI}. Theorem \ref{theorem1} is then proved.

\subsection{Derivation of RW solutions to the 2-LSRI model.}

In this Section, we construct the rational RW solutions to the 2-LSRI model
by starting with matrix elements $m_{s,j}^{(n,k)}$ of tau functions $%
\tau_{n,k}$ \eqref{tau} in the following form:
\begin{equation}  \label{AB-M}
\begin{aligned} m_{s,j}^{(n,k)}=&\mathcal{A}_s\mathcal{B}_jm^{(n,k)},\\
m^{(n,k)}=&\frac{1}{p+q}(-\frac{p-\alpha_1}{q+\alpha_1})^n(-\frac{p-%
\alpha_2}{q+\alpha_2})^ke^{\xi+\eta},\\
\psi_s^{(n,k)}=&\mathcal{A}_s(p-\alpha_1)^{n}(p-\alpha_2)^{k}e^{\xi}, \\
\phi_s^{(n,k)}=&\mathcal{B}_j(q+\alpha_1)^{-n}(q+\alpha_2)^{-k}e^{\eta},
\end{aligned}
\end{equation}
where
\begin{equation}
\begin{aligned}
\mathcal{A}_s=&\frac{\left[\mathcal{T}(p)\partial_{p}\right]^{s}}{s!},
\mathcal{B}_j=\frac{\left[\mathcal{\widehat{T}}(q)\partial_q%
\right]^{j}}{j!}, \\
\xi=&\frac{1}{p-\alpha_2}s+\frac{1}{p-\alpha_1}r+px_1+p^2x_2+\xi_0,\\
\eta=&\frac{1}{q+\alpha_2}s+\frac{1}{q+\alpha_1}r+qx_1-q^2x_2+\eta_0,
\end{aligned}
\end{equation}
and $\mathcal{T}(p)$ and $\mathcal{\widehat{T}}(q)$ are arbitrary functions
of $p$ and $q$, respectively. It is easy to see that these functions also
satisfy the differential and difference relations \eqref{eq:m}, thus $%
\tau_{n,k}=\det(m_{s,j}^{(n,k)})$ with \eqref{AB-M} satisfies the bilinear
equations \eqref{Bi-2-LSRI}.

Then, we constrain the tau functions $\tau_{n,k}=\det(m_{s,j}^{(n,k)})$ with %
\eqref{AB-M} satisfying the dimension reduction condition \eqref{dimension}.
One can directly obtain that:
\begin{equation}  \label{TX}
\begin{aligned}
\mathcal{L}_0m_{s,j}^{(n,k)}=&\mathcal{A}_s\mathcal{B}_jm^{(n,k)}=%
\mathcal{A}_s\mathcal{B}_j\left(\mathcal{Q}(p)+\mathcal{\widehat{Q}}(q)%
\right)m^{(n,k)}, \end{aligned}
\end{equation}
where
\begin{equation}  \label{root-QQ}
\begin{aligned}
&\mathcal{Q}(p)=\frac{\delta_1\rho_1^2}{p-ik_1}+\frac{\delta_2%
\rho_2^2}{p-ik_2}+ip^2,\\
&\mathcal{\widehat{Q}}(q)=\frac{\delta_1\rho_1^2}{q+ik_1}+\frac{\delta_2%
\rho_2^2}{q+ik_2}-iq^2. \end{aligned}
\end{equation}
Following the works \cite{orw-1,orw-2}, if functions $\mathcal{T}(p)$ and $%
\mathcal{\widehat{T}}(q)$ are taken as
\begin{equation}  \label{TT-X1}
\begin{aligned} \mathcal{T}(p)=\frac{\mathcal{W}(p)}{\mathcal{W}^{'}(p)},
\mathcal{\widehat{T}}(q)=\frac{\mathcal{\widehat{W}}(q)}{\mathcal{%
\widehat{W}}^{'}(q)}, \end{aligned}
\end{equation}
we get $\mathcal{T}(p)\partial_p=\partial_{\ln\mathcal{W}},\mathcal{\widehat{%
T}}(q)\partial_q=\partial_{\ln\mathcal{\widehat{W}}}$. Upon the Leibnitz
rules, Eq. \eqref{TX} is rewritten as
\begin{equation}  \label{L-dimen-red}
\begin{aligned}
\mathcal{L}_0m_{s,j}^{(n,k)}=&\sum\limits_{\mu=0}^{s}\frac{1}{\mu!}\left[(%
\mathcal{T}(p)\partial_p)^{\mu}\mathcal{Q}(p)\right]m_{s-\mu,j}^{(n,k)}+
\sum\limits_{l=0}^{j}\frac{1}{l!}\left[(\mathcal{\widehat{T}}(q)%
\partial_q)^{\ell}\mathcal{\widehat{Q}}(q)\right]m_{s,j-l}^{(n,k)}\\
=&\sum\limits_{\mu=0}^{s}\frac{1}{\mu!}\left[\mathcal{W}(p)+(-1)^\mu%
\frac{1}{\mathcal{W}(p)}\right]m_{s-\mu,j}^{(n,k)}+
\sum\limits_{l=0}^{j}\frac{1}{l!}\left[\mathcal{\widehat{W}}(q)+(-1)^l%
\frac{1}{\mathcal{\widehat{W}}(q)}\right]m_{s,j-l}^{(n,k)}. \end{aligned}
\end{equation}

As in Ref. \cite{orw-2}, selecting proper forms of functions $\mathcal{T}(p),%
\mathcal{\widehat{T}}(q)$ can give rise to the coefficients of certain
indices on the right hand side of the above equation vanishing at some
values of $p,q$. To this end, we will choose $p=p_0,q=q_0$ as the roots of
the following algebraic equation:
\begin{equation}  \label{alge-A}
\begin{aligned} \frac{\partial \mathcal{Q}}{\partial p}=0, \frac{\partial
\mathcal{\widehat{Q}}}{\partial q}=0, \end{aligned}
\end{equation}
namely,
\begin{equation}  \label{alge-AXA}
\begin{aligned} \mathcal{Q}^{'}(p_0)=0, \mathcal{\widehat{Q}}^{'}(q_0)=0.
\end{aligned}
\end{equation}
Here $p_0,q_0$ are not pure imaginary, and $p_{0R}>0,q_{0R}>0$.

\textit{\textbf{(1) When $p_0,q_0$ are simple roots of the algebraic
equation \eqref{alge-A}}}, we impose functions $\mathcal{T}(p),\mathcal{%
\widehat{T}}(q)$ meeting the following condition:
\begin{equation}  \label{con-1root}
\begin{aligned}
\left(\mathcal{T}(p)\partial_p\right)^2\mathcal{Q}(p)=\mathcal{Q}(p),\left(%
\mathcal{\widehat{T}}(q)\partial_q\right)^2\mathcal{\widehat{Q}}(q)=%
\mathcal{\widehat{Q}}(q). \end{aligned}
\end{equation}
From Eq. \eqref{TT-X1}, the above condition is rewritten as
\begin{equation}
\begin{aligned}
\partial^2_{\ln{\mathcal{W}}}\mathcal{Q}(p)=\mathcal{Q}(p),\partial^2_{\ln{%
\mathcal{\widehat{W}}}}\mathcal{\widehat{Q}}(q)=\mathcal{\widehat{Q}}(q),
\end{aligned}
\end{equation}
With Eq. \eqref{alge-A} and the scaling $\mathcal{W}(p_0)=1,\mathcal{%
\widehat{W}}(q_0)=1$, we obtain:
\begin{equation}
\begin{aligned}
\mathcal{Q}(p)=\frac{1}{2}Q(p_0)\left(\mathcal{W}(p)+\frac{1}{%
\mathcal{W}(p)}\right),
\mathcal{\widehat{Q}}(q)=\frac{1}{2}\mathcal{\widehat{Q}}(q_0)\left(%
\mathcal{\widehat{W}}(q)+\frac{1}{\mathcal{\widehat{W}}(q)}\right)
\end{aligned}
\end{equation}
and
\begin{equation}  \label{C1-W}
\begin{aligned}
\mathcal{W}(p)=\frac{\mathcal{Q}(p)\pm\sqrt{\mathcal{Q}^2(p)-%
\mathcal{Q}^2(p_0)}}{\mathcal{Q}(p_0)},
\mathcal{\widehat{W}}(q)=\frac{\mathcal{\widehat{Q}}(q)\pm\sqrt{\mathcal{%
\widehat{Q}}^2(q)-\mathcal{\widehat{Q}}^2(q_0)}}{\mathcal{%
\widehat{Q}}(q_0)}. \end{aligned}
\end{equation}
From Eq. \eqref{TT-X1}, the explicit forms of $\mathcal{T}(p),\mathcal{%
\widehat{T}}(q)$ are given through $\mathcal{Q}(p),\mathcal{\widehat{Q}}(q)$
as:
\begin{equation}  \label{XM-T}
\begin{aligned}
\mathcal{T}(p)=&\sqrt{\frac{\mathcal{Q}^2(p)-\mathcal{Q}^2(p_0)}{%
\mathcal{Q}^{'2}(p)}},
\mathcal{\widehat{T}}(q)=&\sqrt{\frac{\mathcal{\widehat{Q}}^2(q)-\mathcal{%
\widehat{Q}}^2(q_0)}{\mathcal{\widehat{Q}}^{'2}(q)}}. \end{aligned}
\end{equation}
The Eq. \eqref{L-dimen-red} becomes
\begin{equation}  \label{L-dimen-red1}
\begin{aligned}
\left.\mathcal{L}_0m_{s,j}^{(n,k)}\right|_{p=p_0,q=q_0}=&\left.%
\mathcal{Q}(p_0)\sum\limits_{\mu=0}^{s}\frac{1}{\mu!}m_{s-\mu,j}^{(n,k)}%
\right|_{p=p_0,q=q_0}+
\left.\mathcal{\widehat{Q}}(q_0)\sum\limits_{l=0}^{j}%
\frac{1}{l!}m_{s,j-l}^{(n,k)}\right|_{p=p_0,q=q_0}. \end{aligned}
\end{equation}
Then, for the restricted indices of the determinants \eqref{tau} of the tau
functions:
\begin{equation}  \label{1-r-m}
\begin{aligned} \tau_{n,k}=\det\limits_{1\leq s,j\leq
N}\left(m_{2s-1,2j-1}^{(n,k)}\right), \end{aligned}
\end{equation}
upon the relation \eqref{L-dimen-red}, the above tau functions satisfy the
following relation:
\begin{equation}
\begin{aligned}
\mathcal{L}_0\tau_{n,k}=\left[\mathcal{Q}(p_0)+\mathcal{\widehat{Q}}(q_0)%
\right]N\tau_{n,k}, \end{aligned}
\end{equation}
which is nothing but the dimension reduction condition \eqref{dimension}.

\textit{\textbf{(2) When $p_0^{({\overline{\alpha}})},q_0^{({\overline{\alpha%
}})} ({\overline{\alpha}}=1,2, p_0^{(1)}\neq\pm p_0^{(2)})$ are two simple
roots of the algebraic equation \eqref{alge-A}}}, then the tau functions are
expressed by the following $2\times2$ block determinants:
\begin{equation}  \label{tau-2X2}
\begin{aligned} \tau_{n,k}=\det\left(\begin{matrix}
\tau_{n,k}^{[1,1]}&\tau_{n,k}^{[1,2]}\\
\tau_{n,k}^{[2,1]}&\tau_{n,k}^{[2,2]} \end{matrix}\right), \end{aligned}
\end{equation}
where
\begin{equation}  \label{2-r-m}
\begin{aligned}
\tau_{n,k}^{[{\overline{\alpha}},{\overline{\beta}}]}=\mathrm{mat}_{1\leq
s\leq N_{\overline{\alpha}},1\leq j\leq
N_{\overline{\beta}}}\left(\left.m_{2s-1,2j-1}^{(n,k)}\right|_{p=p_0^{({%
\overline{\alpha}})},q=q_0^{({\overline{\beta}})}}\right),
1\leq{\overline{\alpha}},{\overline{\beta}}\leq2, \end{aligned}
\end{equation}
and $m_{s,j}^{(n,k)}$ are given by Eq. \eqref{AB-M} with $[\mathcal{T}(p),%
\mathcal{\widehat{T}}(q),\xi_0,\eta_0]$ replaced by $[\mathcal{T}^{[{%
\overline{\alpha}}]}(p),\mathcal{\widehat{T}}^{[{\overline{\beta}}%
]}(q),\xi_{0,{\overline{\alpha}}},\eta_{0,{\overline{\beta}}}]$, here the
functions $\mathcal{T}^{[{\overline{\alpha}}]}(p),\mathcal{\widehat{T}}^{[{%
\overline{\beta}}]}(q)$ are provided by \eqref{XM-T} with $p_0,q_0$ replaced
by $p_0^{({\overline{\alpha}})}$ and $q_0^{({\overline{\beta}})}$ for ${%
\overline{\alpha}},{\overline{\beta}}=1,2$, respectively. These tau
functions \eqref{tau-2X2} also satisfy the bilinear equations \eqref{KPeq},
which can be proved by the same method given in Appendix C of Ref. \cite%
{orw-2}, thus the proof is omitted here.

For this $2\times2$ block determinant forms of tau functions \eqref{tau-2X2}%
, the contiguity relation \eqref{L-dimen-red} is written as
\begin{equation}  \label{L-dimen-red2}
\begin{aligned}
\left.\mathcal{L}_0m_{s,j}^{(n,k)}\right|_{p=p_0^{({\overline{%
\alpha}})},q=q_0^{({\overline{\beta}})}}=&\left.\mathcal{Q}(p_0^{({%
\overline{\alpha}})})\sum\limits_{\mu=0}^{s}\frac{1}{\mu!}m_{s-%
\mu,j}^{(n,k)}\right|_{p=p_0^{({\overline{\alpha}})},q=q_0^{({\overline{%
\beta}})}}
+\left.\mathcal{\widehat{Q}}(q_0^{({\overline{\beta}})})\sum%
\limits_{l=0}^{j}\frac{1}{l!}m_{s,j-l}^{(n,k)}\right|_{p=p_0^{({\overline{%
\alpha}})},q=q_0^{({\overline{\beta}})}}. \end{aligned}
\end{equation}
Similar to the results reported in Ref. \cite{NLS-2}, this contiguity
relation can further yield:
\begin{equation}
\begin{aligned}
\mathcal{L}_0\tau_{n,k}=\left\{\left[\mathcal{Q}(p_0^{(1)})+\mathcal{%
\widehat{Q}}(q_0^{(1)})\right]N_1+\left[\mathcal{Q}(p_0^{(2)})+\mathcal{%
\widehat{Q}}(q_0^{(2)})\right]N_2\right\}\tau_{n,k} \end{aligned}
\end{equation}
namely, the tau functions \eqref{tau-2X2} expressed by $2\times2$ block
determinant also satisfy the dimension reduction condition \eqref{dimension}.

\textit{\textbf{(3) When $p_0,q_0$ are double roots of the algebraic
equation \eqref{alge-A}}}, namely,
\begin{equation}  \label{condition-root-c3}
\begin{aligned} \left.\frac{\partial Q}{\partial
p}\right|_{p=p_0}=\left.\frac{\partial^2 Q}{\partial
p^2}\right|_{p=p_0}=0,\, \left.\frac{\partial \widehat{Q}}{\partial
q}\right|_{q=q_0}=\left.\frac{\partial^2 \widehat{Q}}{\partial
q^2}\right|_{q=q_0}=0, \end{aligned}
\end{equation}
$\mathcal{T}(p),\mathcal{\widehat{T}}(p)$ have the same forms as in Eq. %
\eqref{TT-X1}, but they have to further meet the following condition:
\begin{equation}
\begin{aligned}
\left(\mathcal{T}(p)\partial_p\right)^3\mathcal{Q}(p)=\mathcal{Q}(p),\left(%
\mathcal{\widehat{T}}(q)\partial_q\right)^3\mathcal{\widehat{Q}}(p)=%
\mathcal{\widehat{Q}}(q), \end{aligned}
\end{equation}
namely,
\begin{equation}
\begin{aligned} \partial^{3}_{\ln\mathcal{W}}\mathcal{Q}(p)=\mathcal{Q}(p),
\partial^{3}_{\ln\mathcal{\widehat{W}}}\mathcal{\widehat{Q}}(q)=\mathcal{%
\widehat{Q}}(q). \end{aligned}
\end{equation}
Scaling $\mathcal{W}(p_0)=\mathcal{\widehat{W}}(q_0)=1$, a solution to above
equation under conditions \eqref{condition-root-c3} is
\begin{equation}  \label{3c-QW}
\begin{aligned}
\mathcal{Q}(p)=\frac{\mathcal{Q}(p_0)}{3}\left(\mathcal{W}(p)+\frac{2}{%
\sqrt{\mathcal{W}(p)}}\cos\left[\frac{\sqrt{3}}{2}\ln
\mathcal{W}(p)\right]\right),\\
\mathcal{\widehat{Q}}(q)=\frac{\mathcal{\widehat{Q}}(q_0)}{3}\left(\mathcal{%
\widehat{W}}(q)+\frac{2}{\sqrt{\mathcal{\widehat{W}}(q)}}\cos\left[\frac{%
\sqrt{3}}{2}\ln \mathcal{\widehat{W}}(q)\right]\right). \end{aligned}
\end{equation}
With $\mathcal{Q}(p),\mathcal{\widehat{Q}}(q)$ being of this form, the
following tau functions expressed by block determinants:
\begin{equation}  \label{tau-2X2-3C}
\begin{aligned} \tau_{n,k}=\det\left(\begin{matrix}
\tau_{n,k}^{[1,1]}&\tau_{n,k}^{[1,2]}\\
\tau_{n,k}^{[2,1]}&\tau_{n,k}^{[2,2]} \end{matrix}\right), \end{aligned}
\end{equation}
where
\begin{equation}  \label{3-r-m}
\begin{aligned} \tau_{n,k}^{[\alpha,\beta]}=\mathrm{mat}_{1\leq s\leq
\widehat{N}_\alpha,1\leq j\leq
\widehat{N}_\beta}\left(\left.m_{3s-\alpha,3j-\beta}^{(n,k)}%
\right|_{p=p_0^{(\alpha)},q=q_0^{(\beta)}}\right), 1\leq\alpha,\beta\leq2,
\end{aligned}
\end{equation}
also satisfy the bilinear equations \eqref{KPeq}. This result can be proved
using the method given in Appendix C of Ref. \cite{orw-2}, thus we omit here
its proof. Using the above relations, the Eq. \eqref{L-dimen-red} becomes :
\begin{equation}  \label{L-dimen-red3}
\begin{aligned}
\left.\mathcal{L}_0m_{s,j}^{(n,k)}\right|_{p=p_0,q=q_0}=&\left.%
\mathcal{Q}(p_0)\sum\limits_{{\scriptsize \begin{matrix} \mu=0 \\
\mu\equiv0({\bf mod}3) \end{matrix}}
}^{s}\frac{1}{\mu!}m_{s-\mu,j}^{(n,k)}\right|_{p=p_0,q=q_0}
+\left.\mathcal{\widehat{Q}}(q_0)\sum\limits_{{\scriptsize \begin{matrix}
l=0 \\ l\equiv0({\bf mod}3) \end{matrix}}
}^{j}\frac{1}{l!}m_{s,j-l}^{(n,k)}\right|_{p=p_0,q=q_0}. \end{aligned}
\end{equation}
Upon this contiguity relation, then the tau function \eqref{tau-2X2}
expressed by $2\times2$ block determinant also satisfy the dimension
reduction condition \eqref{dimension}.

Finally, we impose the complex conjugacy condition
\begin{equation}  \label{complex}
\begin{aligned} \tau_{n,k}^*=\tau_{-n,-k}. \end{aligned}
\end{equation}
For this purpose, we first take $\eta_0=\xi_0^*$ in Eq. \eqref{1-r-m} for a
simple root, and $\eta_{0,\beta}=\xi_{0,\alpha}^*$ in Eq. \eqref{2-r-m} and
Eq. \eqref{3-r-m} for two simple roots and a double root. Then, by applying
the change of independent variables \eqref{variable}, it is easy to find
that $\mathcal{\widehat{Q}}$ will be the conjugate of $\mathcal{Q}$ in Eq. %
\eqref{root-QQ} if $q=p^*$, thus $q_0^*=p_0$. When $p_0,q_0$ are single
simple roots of Eq. \eqref{alge-A}, then
\begin{equation}
\begin{aligned}
\left.m_{j,s}^{(-n,-k)}\right|_{p=p_0,q=p_0^*}=\left.\left[m_{s,j}^{(n,k)}%
\right]^*\right|_{p=p_0,q=p_0^*}, \end{aligned}
\end{equation}
thus the complex conjugacy condition \eqref{complex} is realized. If $%
p_0^{[\alpha]},q_0^{[\beta]}$ are two simple single roots of Eq. %
\eqref{alge-A}, since $q_0^{(\alpha)}=p_0^{(\alpha)*}$, thus
\begin{equation}
\begin{aligned}
\left.m_{j,s}^{(-n,-k)}\right|_{p=p_0^{(\alpha)},q=p_0^{(\beta)*}}=\left.%
\left[m_{s,j}^{(n,k)}\right]^*\right|_{p=p_0^{(\beta)},q=p_0^{(\alpha)*}},
\end{aligned}
\end{equation}
which can further imply:
\begin{equation}
\begin{aligned} \tau_{n,k}^{[\alpha,\beta]}=\tau_{-n,-k}^{[\beta,\alpha]*},
\end{aligned}
\end{equation}
then the complex conjugacy condition \eqref{complex} is also satisfied. When
$p_0$ is a double root of Eq. \eqref{alge-A}, the complex conjugacy
condition \eqref{complex} can also be proved in a similar way.

To obtain the solutions in operator differential forms in Theorem \ref%
{theorem-RW-1}--\ref{theorem-RW-3}, we take
\begin{equation}
\begin{aligned}
\xi_0=\sum\limits_{r=1}^{\infty}\widehat{a}_{r}\ln^r\mathcal{W}(p),
\end{aligned}
\end{equation}
for a simple root $p_0$ of Eq. \eqref{alge-A}, where $\mathcal{W}(p)$ is
given by Eq. \eqref{C1-W}, and
\begin{equation}
\begin{aligned}
\xi_{0,\alpha}=\sum\limits_{r=1}^{\infty}\widehat{a}_{r,\alpha}\ln^r%
\mathcal{W}^{(\alpha)}(p),\alpha=1,2 \end{aligned}
\end{equation}
for two simple roots $p_0^{(\alpha)}$ and $p_0^{(\beta)}$ of Eq. %
\eqref{alge-A}, where $\mathcal{W}^{(\alpha)}(p)$ is defined in Eq. %
\eqref{C1-W} with $p_0$ being replaced by $p_0^{(\alpha)}$, and
\begin{equation}
\begin{aligned}
\xi_{0,\alpha}=\sum\limits_{r=1}^{\infty}\widehat{a}_{r,\alpha}\ln^r%
\mathcal{W}(p),\alpha=1,2 \end{aligned}
\end{equation}
for a double root of Eq. \eqref{alge-A}, where $\mathcal{W}(p)$ is given by
Eq. \eqref{3c-QW}. Here $\widehat{a}_r,\widehat{a}_{r,\alpha}$ are arbitrary
complex parameters. Then we can obtain the solutions in differential
operator forms in Theorem \ref{theorem-RW-1}--\ref{theorem-RW-3}.

The last content of this Section is to convert the solutions given through
differential operator forms to Schur polynomials in Theorems \ref%
{theorem-RW-1}, \ref{theorem-RW-2}, \ref{theorem-RW-3}. For this purpose, we
use the following generator $\mathcal{G}$ of differential operators $\left[%
\mathcal{T}(p)\partial_{p}\right]^s$ and $\left[\mathcal{\widehat{T}}%
(q)\partial_{q}\right]^j$ introduced in Ref. \cite{NLS-2},
\begin{equation}
\begin{aligned}
\mathcal{G}=\sum\limits_{s=0}^{\infty}\sum\limits_{j=0}^{\infty}\frac{%
\kappa^s}{s!}\frac{\lambda^j}{j!}\left[\mathcal{T}(p)\partial_{p}\right]^s%
\left[\mathcal{\widehat{T}}(q)\partial_{q}\right]^j. \end{aligned}
\end{equation}
From the relations between $\mathcal{T},\mathcal{\widehat{T}}$ and $\mathcal{%
W},\mathcal{\widehat{W}}$ in Eq. \eqref{TT-X1}, the generator $\mathcal{G}$
can also be written as
\begin{equation}
\begin{aligned}
\mathcal{G}=\sum\limits_{s=0}^{\infty}\sum\limits_{j=0}^{\infty}\frac{%
\kappa^s}{s!}\frac{\lambda^j}{j!}\left[\partial_{\mathcal{W}(p)}\right]^s%
\left[\partial_{\mathcal{\widehat{W}}(q)}\right]^j
=\exp\left(\kappa\partial_{\ln\mathcal{W}(p)}+\lambda\partial_{\ln\mathcal{%
\widehat{W}}(q)}\right), \end{aligned}
\end{equation}
The operator $\mathcal{G}$ acting on an arbitrary function $\mathcal{F}(%
\mathcal{W},\mathcal{\widehat{W}})$ can result in the following relation:
\begin{equation}
\begin{aligned}
\mathcal{G}\mathcal{F}(\mathcal{W},\mathcal{\widehat{W}})=\mathcal{F}%
\left(e^{\kappa}\mathcal{W},e^{\lambda}\mathcal{\widehat{W}}\right).
\end{aligned}
\end{equation}
Since the parameter $p$ and $q$ are related to $\mathcal{W}$ and $\mathcal{%
\widehat{W}}$, respectively, thus one can regard $p$ as a function of $%
\mathcal{W}$, and $q$ as a function of $\mathcal{\widehat{W}}$, namely, $p=p(%
\mathcal{W}),q=q(\mathcal{\widehat{W}})$. Furthermore, $\mathcal{W}(p_0)=%
\mathcal{\widehat{W}}(q_0)=1$.

For $m^{(n,k)}$ in Eq. \eqref{rw1-m} of Theorem \ref{theorem-RW-1},
\begin{equation}  \label{Sch-1C}
\begin{aligned}
&\left.\frac{1}{m^{(n,k)}}\mathcal{G}m^{(n,k)}\right|_{p=p_0,q=q_0}\\
=&\frac{p_0+q_0}{p(\kappa)+q(\lambda)}\left(\frac{p(\kappa)-ik_1}{p_0-ik_1}%
\right)^n\left(\frac{q(\lambda)+ik_1}{q_0+ik_1}\right)^{-n}
\left(\frac{p(\kappa)-ik_2}{p_0-ik_2}\right)^k\left(\frac{q(%
\lambda)+ik_2}{q_0+ik_2}\right)^{-k}\\
&\exp\left(\sum\limits_{r=1}^{\infty}\widehat{a}_r\kappa^r+\widehat{a}_r^*%
\lambda^r\right)
\times\exp\left[(p(\kappa)-p_0+q(\lambda)-q_0)x-i(p^2(\kappa)-p_0^2-q^2(%
\lambda)+q^2_0)t\right]. \end{aligned}
\end{equation}
Here we have set $q=p^*$ in Eq. \eqref{rw1-m}. After expanding the right
side of the above equation into power series of $\kappa$ and $\lambda$, the
first term can be written into the following form \cite{orw-1,orw-2}:
\begin{equation}
\begin{aligned}
\frac{p_0+q_0}{p(\kappa)+q(\lambda)}=\sum\limits_{v=0}^{\infty}\left(\frac{|%
\lambda_1|^2}{(p_0+q_0)^2}\kappa
v\right)^v\exp\left(\sum\limits_{r=1}^{\infty}(vs_r-b_r)\kappa^r+
(vs_r^*-b_r^*)\lambda^r\right), \end{aligned}
\end{equation}
where $\lambda_1$ is given by Eq. \eqref{rw1-abl}, and $b_r$ is given by the
Taylor coefficient of $\kappa^r$ in the expansion of
\begin{equation}
\begin{aligned}
\ln\left[\frac{p(\kappa)+q_0}{p_0+q_0}\right]=\sum\limits_{r=1}^{\infty}b_r%
\kappa^r, \end{aligned}
\end{equation}
and $s_r$ is defined in Eq. \eqref{rw1-s}. The rest term can be written as
\begin{equation}
\begin{aligned}
\exp&\left\{\sum\limits_{r=1}^{\infty}\kappa^r\left[\lambda_rx-i\beta_r
t+n\theta_r^{(1)}+k\theta_r^{(1)}\right]
+\sum\limits_{r=1}^{\infty}\lambda^r\left[\lambda^*_rx+i\beta^*_r
t\right.\right.\\ &\left.\left.+n\theta_r^{(1)}
+k\theta_r^{(1)}\right]+\sum\limits_{r=1}^{\infty}\left(\widehat{a}_r%
\kappa^r+\widehat{a}_r^*\kappa^{r}\right)\right\}. \end{aligned}
\end{equation}
Combining these expansions, then Eq. \eqref{Sch-1C} becomes
\begin{equation}  \label{expand-m1}
\begin{aligned}
&\left.\frac{1}{m^{(n,k)}}\mathcal{G}m^{(n,k)}\right|_{p=p_0,q=q_0}\\
=&\sum\limits_{v=0}^{\infty}\left(\frac{|\lambda_1|^2}{(p_0+q_0)^2}\kappa%
\lambda\right)^v\exp\left(\sum\limits_{r=1}^{\infty}(x^{+}_r+vs_r)\kappa_r
+\sum\limits_{r=1}^{\infty}(x_r^{-}+vs^*)\lambda^r\right), \end{aligned}
\end{equation}
where $x_r^{\pm}(n,k)$ are as given by Eq. \eqref{rw1-x} with
\begin{equation}
\begin{aligned} a_r=\widehat{a}_r-b_r. \end{aligned}
\end{equation}
Taking the coefficients of $\kappa^s\lambda^j$ on both sides of Eq. %
\eqref{expand-m1}, we get
\begin{equation}  \label{m1-m1}
\begin{aligned}
\frac{m_{s,j}^{(n,k)}}{m^{(n,k)}|_{p=p_0,q=q_0}}=\sum\limits_{v=0}^{%
\mathrm{min}(s,j)}\left(\frac{|\lambda_1|^2}{(p_0+q_0)^2}\right)^v%
\mathbf{S}_{s-v}(\mathbf{x}^{+}(n,k)+v\mathbf{s})
\mathbf{S}_{j-v}(\mathbf{x}^{-}(n,k)+v\mathbf{s}^*), \end{aligned}
\end{equation}
where $m_{s,j}^{(n,k)}$ is the matrix element given by Eq. \eqref{rw1-m}.
The right side of the above equation is the $\widehat{m}_{s,j}^{(n,k)}$
defined in Eq. \eqref{rw1-cm}. Furthermore, from Eq. \eqref{m1-m1} we can
obtain the following relation:
\begin{equation}
\begin{aligned} \det\limits_{1\leq s,j\leq
N}\left(m_{2s-1,2j-1}^{(n,k)}\right)=\widehat{H}\det\limits_{1\leq s,j\leq
N}\left(\widehat{m}_{2s-1,2j-1}^{(n,k)}\right) \end{aligned}
\end{equation}
where $\widehat{H}=\left(m^{(n,k)}|_{p=p_0,q=q_0}\right)^N$. This relation
indicates that the solutions given through differential operator form and
through Schur polynomials in Theorem \ref{theorem-RW-1} are equivalent.

Similarly, we can also transform the solutions expressed by differential
operator into Schur polynomials in this way. This completes the proof of
Theorems \ref{theorem-RW-1}, \ref{theorem-RW-2}, \ref{theorem-RW-3}.

\section{Conclusion and discussion}\label{conclusion}
In this paper, we have constructed general higher-order
breather and RW solutions of the 2-LSRI model \eqref{2-LSRI} in the form of
determinants, by means of the bilinear KP-hierarchy reduction method. We
first studied the dynamics of the breather solutions. Under particular
restrictions imposef on the parameters, the breather solutions can become
the homoclinic orbits in the 2-LSRI model \eqref{2-LSRI}. It has been shown
that the second-order breather solutions have three different dynamical
behaviours: two breathers, second-order homoclinic orbits, and a mixture of
a breather and a first-order homoclinic orbit. We derived three families of
RW solutions to the 2-LSRI model, which correspond to a simple root, two
simple roots, and double roots of the equation \eqref{root} related to the
dimension reduction condition. They are bounded RW, mixed bounded RWs, and
degradable bounded RWs. The dynamics of these three families of RWs have
been exhibited. The differences between these three families of RWs can be
summarized as follows:

\begin{itemize}
\item The polynomial degree of the tau functions $\tau_{n,k}$ in Theorem \ref%
{theorem-RW-1}--\ref{theorem-RW-3} are $N(N+1)$, $N_1(N_1+1)+N_2(N_2+1)$ and
$2[\widehat{N}_1^2+\widehat{N}_2^2-\widehat{N}_1(\widehat{N}_2-1))$ in both $%
x$ and $t$ variables, respectively, where $N,N_1,N_2$ are positive integers
and $\widehat{N}_1,\widehat{N}_2$ are non-negative integers.

\item The solutions in Theorem \ref{theorem-RW-1}--\ref{theorem-RW-3}
comprise of \, $\frac{N(N+1)}{2}$, \, \, $\frac{N_1(N_1+1)+N_2(N_2+1)}{2}$,
\, \, and ${[\widehat{N}_1^2+\widehat{N}_2^2-\widehat{N}_1(\widehat{N}_2-1)]}
$ fundamental RWs, respectively. The RWs in Theorem \ref{theorem-RW-1} and
Theorem \ref{theorem-RW-3} are bounded states, while they are a mixture of
two different bounded states in Theorem \ref{theorem-RW-2}.
\end{itemize}

We point out that the breathers and RW solutions to the 2-LSRI model have
been derived earlier by using the bilinear method \cite{Chow-Non} and the
Darboux transformation \cite{chen-PRE,geng-CNSNS,YO-Geng1,YO-Geng2}.
Comparing with those previously reported results, the main results obtained
in this paper can be summarized as follows:

\begin{itemize}
\item There were only the first-order breather and first-order RW solutions
given in Ref. \cite{Chow-Non}, while only the first-order RW solutions were
studied in Ref. \cite{chen-PRE}. In the present paper we have constructed
the general higher-order breather and RW solutions in terms of determinants.

\item The higher-order RW solutions given in Refs. \cite%
{geng-CNSNS,YO-Geng1,YO-Geng2} comprise $N(N+1)/2$ fundamental RWs, which
correspond to the family of RW solutions in Theorem \ref{theorem-RW-1} of
the present paper. The general higher-order RW solutions in Theorems \ref%
{theorem-RW-2} and \ref{theorem-RW-3} are new RW solutions to the 2-LSRI
model \eqref{2-LSRI}, which have not been reported before, to the best of
our knowledge.
\end{itemize}

\section*{Acknowledgments}

{\noindent {\small The work was supported by the National Natural Science
Foundation of China (Grants 11871446/12071451), and the Guangdong Basic and Applied Basic Research Foundation (Grant 2022A1515012554),
and the Research and Development Foundation of Hubei University of Science and Technology (Grant BK202302),
 and Israel Science Foundation (grant No. 1286/17).}}

\end{document}